\documentclass{ws-ijmpd}

\begin{document}

\markboth{S. Simi\'c and L.\v C. Popovi\'c}
{Physical Parameters Of The Relativistic Shock Waves In GRBs: The Case Of 30 GRBs}

%%%%%%%%%%%%%%%%%%%%% Publisher's Area please ignore %%%%%%%%%%%%%%%
%
\catchline{}{}{}{}{}
%
%%%%%%%%%%%%%%%%%%%%%%%%%%%%%%%%%%%%%%%%%%%%%%%%%%%%%%%%%%%%%%%%%%%%

\title{PHYSICAL PARAMETERS OF THE RELATIVISTIC SHOCK WAVES IN GRBs: THE CASE OF 30 GRBs}

\author{SA\v SA SIMI\'C}

\address{Faculty of Science, Department of Physics, University of Kragujevac\\
Radoja Domanovi\'ca 12, Kragujevac, Serbia, 34000\\
ssimic@kg.ac.rs}

\author{LUKA \v C. POPOVI\' C}

\address{Astronomical Observatory, Volgina 7\\
Belgrade, Serbia, 11000\\
lpopovic@aob.bg.ac.rs}

\maketitle

\begin{history}
\accepted{18.01.2012 for publication in IJMPD}
\end{history}

\begin{abstract}
Using the modified internal shock wave model we fit the gamma ray burst (GRB) light and spectral curves of 30 GRBs observed
with BATSE. From the best fitting we obtain basic parameters of the relativistic shells which are in good agreement with predictions given earlier. We compare measured GRB parameters with those obtained from the model and discuss connections between them in the frame of the physical processes laying behind GRB events.
\end{abstract}


\section{Introduction}

The quest on resolving a Gamma Ray Burst (GRB) event consists of finding an explanation for several parts e.g. spatial distribution of the event,
afterglow, spectral and light curve, collimation, etc. In the past decade the GRB phenomenon has been thoroughly investigated
both observationally see e.g. [\refcite{Bloom98}]-[\refcite{Ostlin08}]
%\cite{Costa97}, \cite{Groot98}, \cite{Kulkarni98}, \cite{Kulkarni99}, \cite{Andersen99}, \cite{Harrison99}, \cite{Galama99},
%\cite{Price01}, \cite{Yost02}, \cite{Rykoff04}, \cite{Campana06}, \cite{Ostlin08}
and theoretically see e.g. [\refcite{Meszaros97}]-[\refcite{Zitouni08}].
%\cite{Vietri97}, \cite{Waxman97}, \cite{Sari98}, \cite{Huang98}, \cite{Huang99}, \cite{Nakar03}, \cite{Piran05}, \cite{Zitouni08}.
However, a mystery of GRB phenomena lies in its heart, where the central engine ejects material with  relativistic energies and velocities. Due to the high optical depth of the expanding material in the first phase of the GRB event, structural observations of the central engine, that is located near the core of the progenitor, are not possible. The only information that one can obtain from the observations, in the first minute of a GRB event,
is the temporal variability of the $\gamma$-ray light curve. This usually shows strong, but short fluctuations of the energy output with a typical time-scale of the order of milliseconds to a couple of seconds [\refcite{Norris96}]. The numerical simulations of [\refcite{Kobayashi97}] and [\refcite{Nakar02}] for the early phase of the explosion, revealed that  $\gamma $-ray light curve pulses replicate temporal activity of the 'inner engine'. This can give information about the connection between observations and physical processes occurring in the GRB core.

In order to make conclusion about the physical processes in the GRB core one should assume the mechanism of
GRB origin. The mechanism of pulse creation in a GRB light curve is proposed to be connected with mutual interactions
(collisions) between faster and slower spreading shells see e.g. [\refcite{Piran05}], in the so called
internal shock model. Here we accepted slightly modified version of the model with the additional assumption of
the non-zero density environment and also with different treatment of the slower material (shell) accumulated at some
distance from the GRB engine. In a  previous paper [\refcite{Simic07}], where we considered interaction of an
incoming faster shell with dense barrier, we demonstrated that the model is able to reproduce and fit well the observed GRB light curves.

In this paper we apply the model to a sample of 30 GRB light curve pulses observed with Burst and Transient Source Experiment (BATSE) in order to discuss the physical parameters of relativistic shells.

The paper is organized as follow: in \S 2 we give a brief description of the model and considered assumptions; in \S 3 we describe the observational sample and fitting procedure; in \S 4 we discuss our results and finally in \S 5 we outline our conclusions.

\section{The model and assumptions}

Here we describe the physical scenario and some approximations used in the model of GRB light and spectral curves. First, we assume that the GRB engine ejects an amount of relativistically expanding material, that spreads isotropically from the center of explosion. The material is subsequently ejected from time to time depending on the central engine activity.
Here we will not consider the nature of GRB progenitors, i.e. for the model it is not relevant whether a GRB event is originated in the collapse of a massive star [\refcite{Woosley93}] or in the process of merging of two compact objects -- neutron stars or black holes [\refcite{Norris00}], [\refcite{Fryer99}]. The most important for the model is the assumption that the ejected material can mutually interact or interact with the surrounding environment.The ejected material is probably irregular in nature, with different initial parameters (mass $M_{\rm ej}$ and Lorentz factor $\Gamma_{0}$).

During the expansion of the ejected material (closer to the GRB core), a slow moving material is followed by a fast moving one, thus the faster moving material will overtake the slower material and plunge into it. This interaction produces the relativistic shells and shock waves that can accelerate particles to very high speeds. The new formed relativistic shells are probably with different velocities and could collide mutually producing the observed GRB light curve pulses. The observed GRB light curves show vast variety of pulses, ranging from very intense ones to those almost equal to noise, and from symmetrical to highly asymmetrical ones. To model different observational light curve pulse profiles, we consider here the modified standard internal model, assuming that the accumulated matter is able to form a slow moving barrier. This, allowed us to generate more diversity in light curve pulse profiles.

In the shell interaction processes, the mass of a newly created shell is approximately the sum of the two colliding shells and its Lorentz factor is smaller than it was in the faster shell. It could be additionally reduced by surrounding medium or further shell interactions, thus producing the accumulation effect. Depending on the density and width of such created dense barrier, further collision with it produces wider/thinner or higher/lower pulses. Also, considering the energy density of expanding shells and barriers, one can get more symmetrically shaped light curve pulses in case of the shells with high energy and low energy barriers, and high asymmetrical pulse shapes in the opposite case. This is the main mechanism that we use to explain high temporal variability of the observed $\gamma$-ray light curves. From the scenario described above, it follows that the selection of shell parameters can have a random distribution in a given interval of values.

Similar as in [\refcite{Kobayashi97}], we consider that the ejected GRB material is organized in an ultra-relativistic flow of well defined and collimated shells with random initial energy. In contrary to [\refcite{Piran05}] and [\refcite{Kobayashi97}], where null density hypothesis was used, we assume here that the surroundings regions around the GRB central engine consist of at least small number of particles with densities $n_0 > 1\ \rm cm^{-3}$. This allows us to analyze the density of the moving shell and to model the density distribution and shape of the barrier in a specific way (here we use Gaussian function, see Eq. \ref{eq5} further in the text). The assumption of $n_0\neq 0$ around the central engine seems to be generally valid. If there are some kind of ejections from the central engine, one can expect an amount of scattered material distributed to the surrounding region, as e.g. in the collapsar model where a central star is  Wolf-Rayet type. Such star ejects a huge amount of material during its final stage, therefore one can expect non zero density in its environment.

To describe the evolution of a relativistic shell, we adopt a phenomenological model based on the [\refcite{Huang98}] that  presents a system of the first order of differential equations (where the distance $R$, Lorentz factor $\Gamma$ and mass $m_s$ of the shell are included, see [\refcite{Simic07}]):

\begin{equation}
\label{eq1} {\frac{{dR}}{{dt}}} = c\sqrt {\Gamma^{2} - 1} {\left[
{\Gamma + \sqrt {\Gamma^{2} - 1}}  \right]},
\end{equation}

\begin{equation}
\label{eq2} {\frac{{d\Gamma}} {{dm_{s}}} } = - {\frac{{\Gamma^{2}
- 1}}{{M_{\rm ej} + 2(1 - \xi )\Gamma m_{s} + \xi m_{s}}}},
\end{equation}

\begin{equation}
\label{eq3} {\frac{{dm_{s}}} {{dt}}} = 2\pi nm_{p} (1 - \cos
\theta ){\frac{{R^{2}}}{{\Gamma^{3}}}}\left( {3\Gamma
{\frac{{dR}}{{dt}}} - 2R{\frac{{d\Gamma}} {{dt}}}} \right),
\end{equation}
where the parameter $\xi$ takes values from 0 in case of adiabatic expansion to 1 in fully radiative case. $M_{\rm ej}$ and $\theta$ are the initial mass and collimation angle of the shell, $n$ is the number density and $m_p$ is the proton mass. Eqs. (\ref{eq1}) - (\ref{eq3}) are derived for an observer reference frame, and they have to be solved simultaneously, together with the density equation. The initial values of parameters and variables are highly dependent on physical properties of the shocks.

The density of the barrier cerated from accumulating decelerated shells (emitted by the central engine) could be described with  (see [\refcite{Blandford76}]):

\begin{equation}
\frac{n_{2}}{n_{1}}=\frac{\kappa_{2}\gamma_{2} + 1}{\kappa_{2} - 1}
\label{eq4}
\end{equation}
where $n_2$ and $n_1$ are  number densities after and in front of the shock, $\kappa_2$ is the ratio of the specific heat for the shocked fluid and $\gamma_2$  is the Lorentz factor of the shocked fluid.

This equation gives a connection between density of perturbed and unperturbed material. In the case of the ultra-relativistic expansion, the ratio of the specific heats has a constant value of $\kappa=4/3$, then Eq. \ref{eq4}  can be reduced to $\frac{n_{2}}{n_{1}}=4\gamma_{2} + 3$. Also, in the relativistic regime, the Lorentz factor of the shell $\Gamma$ is directly proportional to the Lorentz factor of the shocked particles $\gamma_{2}$ [\refcite{Blandford76}].

In case of the collision of relativistic shells, a slower interacting shell (or barrier in our model) presents a density perturbation in surrounding media for the incoming faster shell. The barrier (or accumulated material) at a distance $R_{c}$ from the central engine should have a distribution of the density along the path of penetration of the faster shell. This can be included in calculation, and here we assume a Gaussian density distribution as:
\begin{equation}
\label{eq5} n = n_{0} \left( {{\frac{{R_{0}}} {{R}}}}
\right)^{s}(4\Gamma + 3)\left( {1 + a \cdot \exp {\left[ { -
\left( {{\frac{{R - R_{c}}} {{b}}}} \right)^{2}} \right]}} \right)
\end{equation}
where $a$ and $b$ describe the Gaussian intensity and width of the barrier and $n_0$ is the density of the surrounding region.
$R_0$ is the initial position of the faster shell, $\Gamma$ is the Lorentz factor of the shell as used in Eqs. (1-3).
Note here that in Eq. \ref{eq5} (as it was mentioned above) $n_{0}>0$, and it is a crucial difference between our and the standard internal shock (IS) model. In the standard IS model, the density
of the interacting shells is fixed by the mass loss rate from
the central engine, the Lorentz factor of the shells and the
distance from the source where the collisions take place. It is
not directly related to the density of the external medium. As
a result, even in the absence of an external medium, prompt
emission will result from shocks taking place in the material
ejected from the source. Indeed there is some evidence of
bursts occurring in very low density environments which have a
prompt emission but no detectable afterglow, but as we mentioned above, one could expect that $n_{0}>0$ (especially close to the accumulated material) in the central engine surroundings.

Generally, one can expect that the ejected material in form of the shell can collide with the ISM which
has a certain density distribution. In such a highly relativistic physical system the relative
motion of charged particles of the ISM can generate an intense magnetic field in the reference
frame of the moving fluid. We calculate the magnetic field in a similar way as in [\refcite{Huang00}], by assuming
that the energy of the magnetic field is a certain fraction, $\xi_{b}$, of the total
energy of the relativistic shell. In the comoving reference frame  the magnetic
field is calculated as:

\begin{equation}
\label{eq8} B^{'} = \sqrt {8\pi \xi _{b} n_{0} \Gamma m_{p}
c^{2}(4\Gamma + 3)\left( {{\frac{{R_{0}}} {{R}}}}
\right)^{s}\left( {1 + a \cdot e^{ - \left( {{\frac{{R - R_{c}}}
{{b}}}} \right)^{2}}} \right)},
\end{equation}
where the variables used in this equation are same as in the Eqs. \ref{eq1}-\ref{eq5}.

The emission mechanism of shock waves is mainly based on the
synchrotron radiation, but for higher energy bands additional
flux may be gained by the Inverse Compton (IC) radiation
[\refcite{Piran05}]. In the first approximation we  neglect the IC radiation.

We calculate the intensity of the radiation emitted by particles in the relativistic shell
using the formulae given by [\refcite{Rybicki79}]. Then the total emitted flux can be calculated
as e.g. in [\refcite{Huang00a}]. Note here that an expanding shell contains relativistic electrons and baryons which
contribute to the synchrotron radiation. However, taking into
account the difference in velocities of these constituents, one
can neglect the contribution of baryons to the total emitted
flux. Then in the comoving reference frame the total flux is given as:

\begin{equation}
\label{eq6} P_{\nu}^{'}=A\cdot \int _{\gamma_{emin}}^{\gamma_{emax}} \gamma_{e}^{-(p+1)}
F(\nu^{'}/\nu_{syn}^{'}) d\gamma_{e}
\end{equation}
{where $A$, $\gamma_{emin}$ and $\gamma_{emax}$ are:}

\begin{equation}
\label{eq7}
A=\frac{\sqrt{3}e^{3}B^{'}}{m_e c^{2}} \frac{m_s}{m_p}; \gamma_{emin}=\xi_e \Gamma \frac{(p - 2)}{(p - 1)} \frac{mp}{me}; \gamma_{emax}\rightarrow \infty;
\end{equation}
 and
$$F(x)=\int_{x}^{\infty} K_{5/3}(x)dx,$$
where, $K_{5/3}$ is the Bessel function of the second order and $\nu_{syn}^{'}=3 \gamma_{e}^{2}e B^{'}/4 \pi m_{e} c$ is
the critical frequency of the synchrotron radiation.

In Eq. \ref{eq6} we neglect the effects of the surface curvature
of an emitting shell, since it has a small influence on the pulse shape.

In the case of ultra relativistic shells the cooling time is much shorter then the dynamical (time of expansion)
(see [\refcite{Molinari07}]). This is particulary interesting in the gamma phase of explosion where shells are interacting with each other.

With the model described above we are able to simulate collisions of relativistic shells in the first phase of a GRB event, which produce the peaks in the light curve [\refcite{Simic07}]. This model has been used to fit the light and energy curves of the  sample of 30 GRBs taken from the BATSE database.

\section{Model vs. observations}

In [\refcite{Simic07}] we demonstrated that the model is able to reproduce (simulate) the observed light
curves of GRBs. Moreover, the model can properly fit the observed light curves. Here we selected 30 GRBs from the BATSE database and fit them with the model in order to find the basic parameters of interacting shells. In this section we describe the selected sample and the fitting procedure.

\subsection{The sample}

From a large BATSE database (3rd channel, $E = 100 - 300$ keV, for the light curve) we select a sample of 30 GRBs
using following criteria:
(i) GRBs have isolated light curves with the clear peak maximum. For proper application of the model we avoid the complex pulse profiles. \footnote{The process of pulse creation is stochastic in nature and may result in a complex pulse
shape, e.g. it is often observed that two or more pulses are superposed;}
(ii) we avoid small pulses because of their low temporal resolution;
(iii) we include in the sample as much as possible different GRBs (long and short lasting,
with different profiles, different intensity and different profile asymmetries).

In Table \ref{tab1} we give a list and basic parameters of selected GRBs. The parameters in the table are (from the first to the last column respectively): the Full Width at Half Maximum of the intensity of light curve pulse (FWHM), the time of peak intensity for the observed pulse ($t_{\rm peak}$), the total duration of the pulse from the beginning to the end of its lower tail ($\Delta t$), the maximal intensity ($J_m$) of the pulse measured in ${\rm erg/cm^{2} s Hz}$, and the asymmetry indicator ($w$) calculated as a ratio of half-halfwidths before and after the maximum.

\begin{table}[ph]
\begin{scriptsize}
\tbl{Parameters of the selected GRB light curves. From the first to the last column the  following parameters are given: Full width at half maximum (FWHM), time of peak intensity $t_{\rm peak}$, duration of the pulse $\Delta t$, pulse intensity ($J_m\rm [erg/cm^{2} s Hz]$), asymmetry indicator $w_s$.}
{\begin{tabular}{@{}cccccc@{}} \toprule
GRB & FWHM [s] &$\Delta t_{\rm peak} [s]$ & $\Delta t [s]$ & $J_m [\times
10^{-27}]$ & $w_s$ \\ \colrule
GRB910629 & 0.55 & 0.4 & 1 & 13 & 0.6 \\
GRB911104 & 0.85 & 0.5 & 1.3 & 13 & 0.5 \\
GRB920715 & 0.5 & 0.45 & 1 & 3.2 & 0.4 \\
GRB920720 & 0.75 & 0.6 & 1.4 & 40 & 0.7 \\
GRB920808 & 0.5 & 0.25 & 0.7 & 11.3 & 0.4 \\
GRB920811 & 0.2 & 0.25 & 0.3 & 3.5 & 0.7 \\
GRB920830 & 2.75 & 1.9 & 6 & 7.8 & 0.6 \\
GRB920912 & 0.4 & 0.58 & 1.2 & 5.4 & 0.7 \\
GRB920924 & 0.4 & 0.4 & 0.8 & 4. & 0.3 \\
GRB921021 & 1.8 & 1 & 4 & 5.1 & 0.4 \\
GRB921207 & 1.4 & 1.4 & 3 & 60. & 0.6 \\
GRB921208 & 1 & 0.85 & 2 & 2.25 & 0.8 \\
GRB921222 & 1 & 0.85 & 2 & 2.25 & 0.8 \\
GRB950129B & 0.6 & 0.65 & 1.22 & 3.7 & 0.7 \\
GRB950211B & 1.5 & 0.95 & 3.5 & 6.5 & 0.4 \\
GRB960111 & 1.2 & 0.5 & 2.6 & 9 & 0.5 \\
GRB960207 & 0.33 & 0.25 & 0.6 & 9.2 & 0.8 \\
GRB960229 & 0.55 & 0.35 & 1 & 5.4 & 0.8 \\
GRB960311 & 0.4 & 0.4 & 0.7 & 6 & 0.6 \\
GRB960409 & 2.4 & 2 & 5 & 10.5 & 0.7 \\
GRB960418 & 0.9 & 0.8 & 1.5 & 4.7 & 0.4 \\
GRB960524 & 0.55 & 0.4 & 1.2 & 6.2 & 0.8 \\
GRB960528 & 2.2 & 0.55 & 4.2 & 4.2 & 0.6 \\
GRB960530 & 5.5 & 3 & 11 & 5.7 & 0.6 \\
GRB960613 & 1.6 & 1.8 & 4 & 6 & 0.3 \\
GRB960617B & 0.7 & 0.55 & 1.1 & 3.3 & 0.8 \\
GRB970424 & 0.4 & 0.3 & 0.7 & 3.3 & 0.5 \\
GRB991105 & 0.65 & 0.45 & 1.6 & 6 & 0.4 \\
GRB991213 & 1.4 & 1 & 2.4 & 4 & 0.4 \\
GRB000107 & 0.5 & 0.45 & 0.7 & 4.7 & 1. \\ \botrule
\end{tabular} \label{tab1}}
\end{scriptsize}
\end{table}

\begin{figure*}
\centering
\includegraphics[width=6cm]{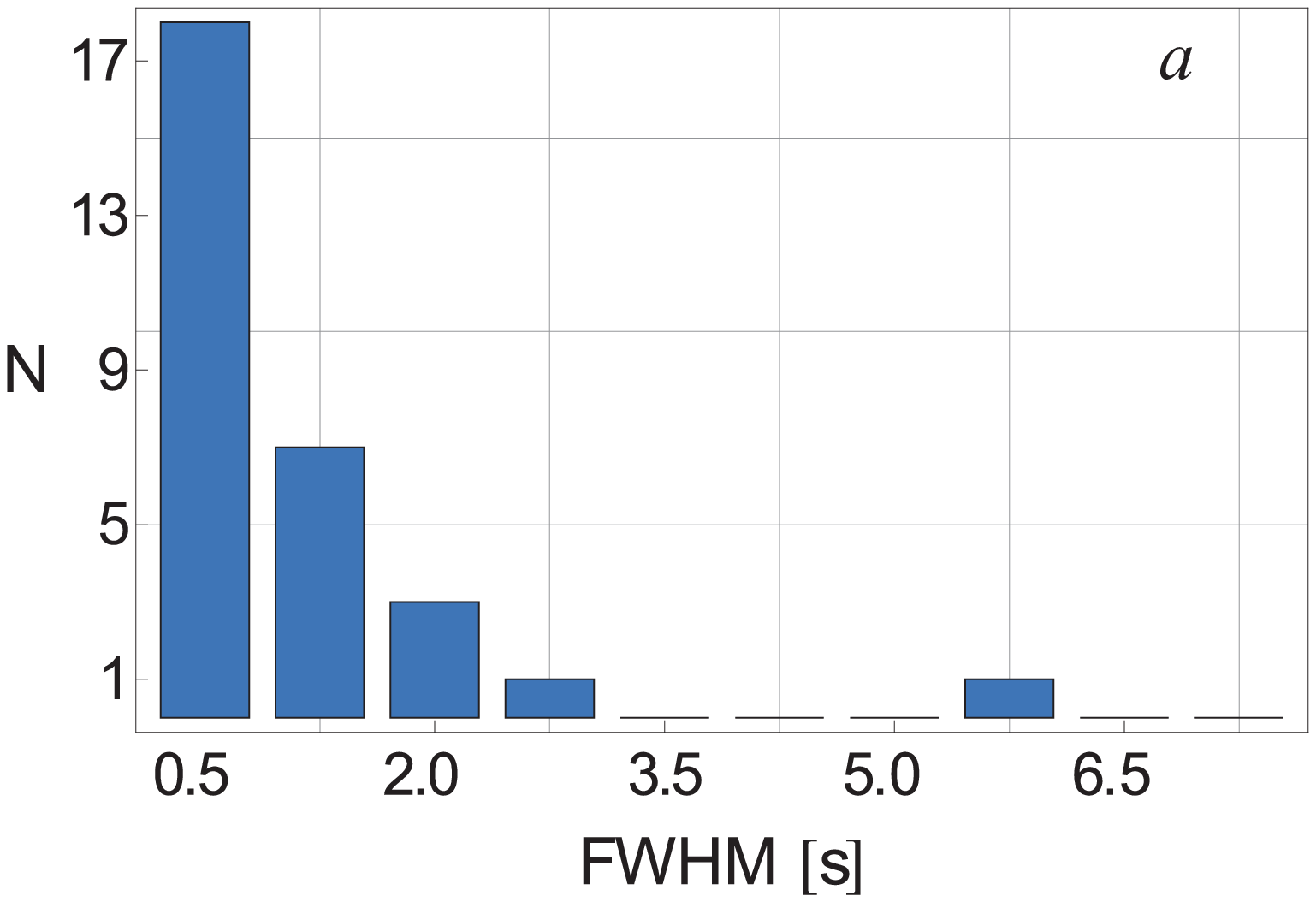}
\includegraphics[width=6cm]{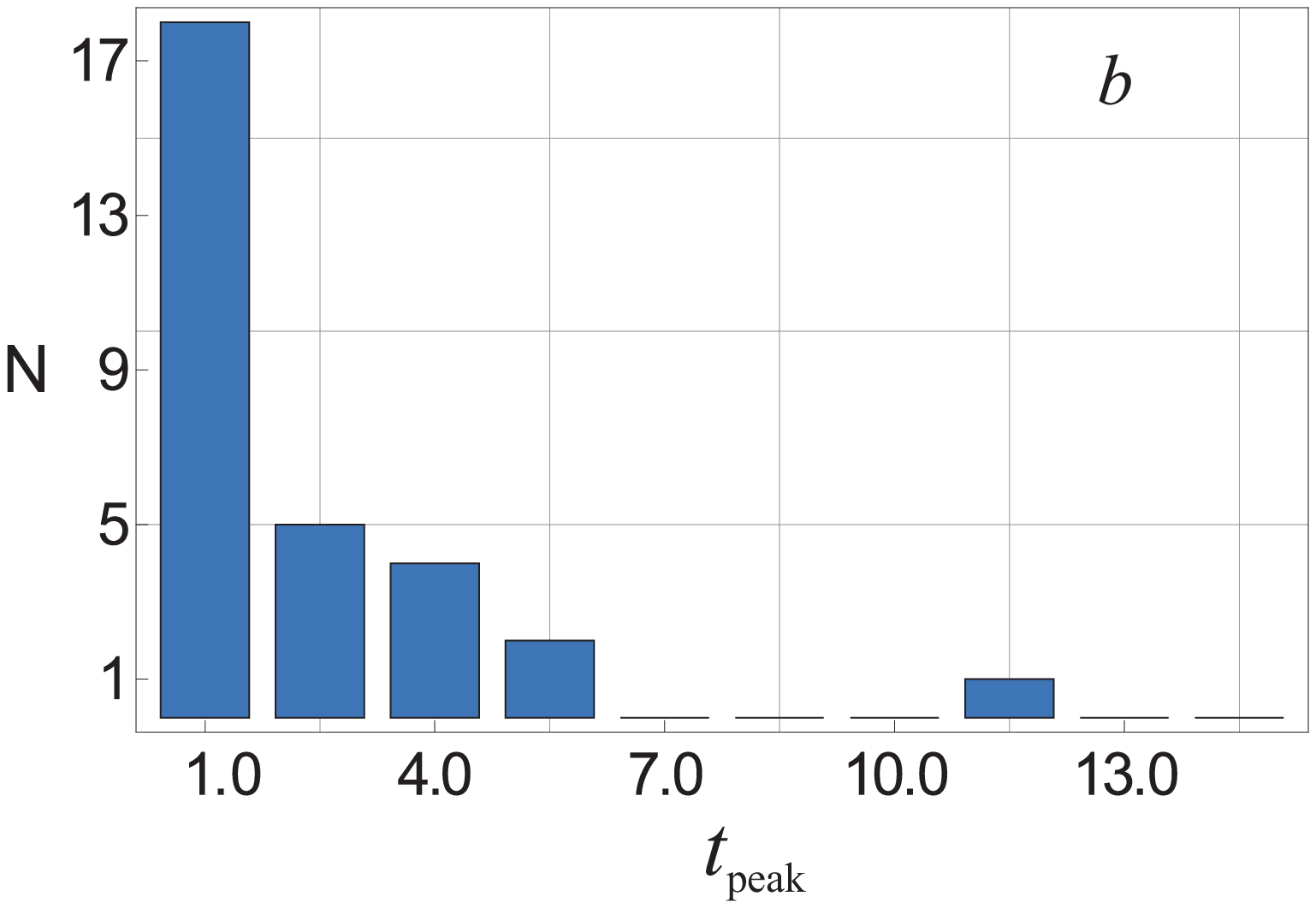}
\includegraphics[width=6cm]{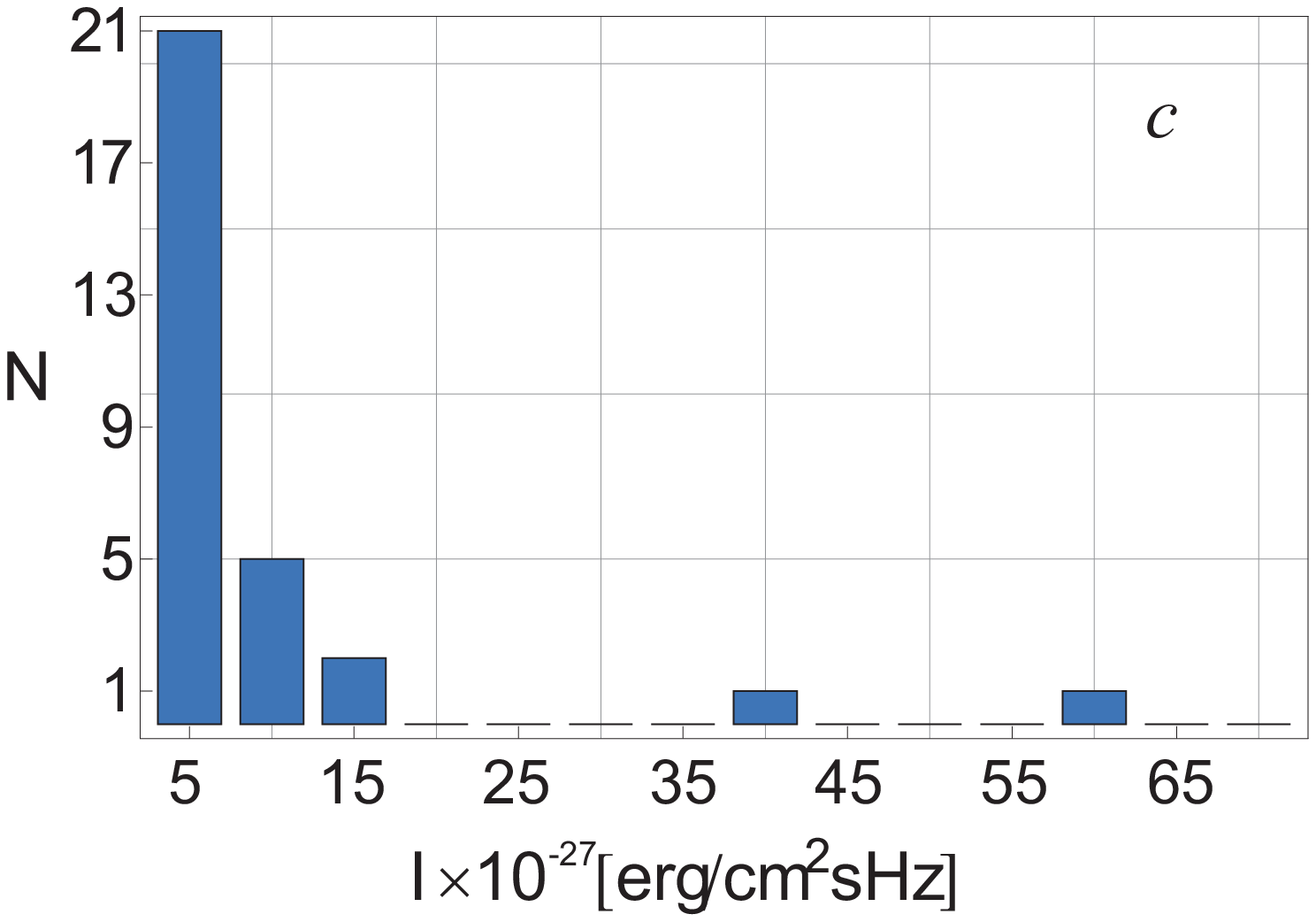}
\includegraphics[width=6cm]{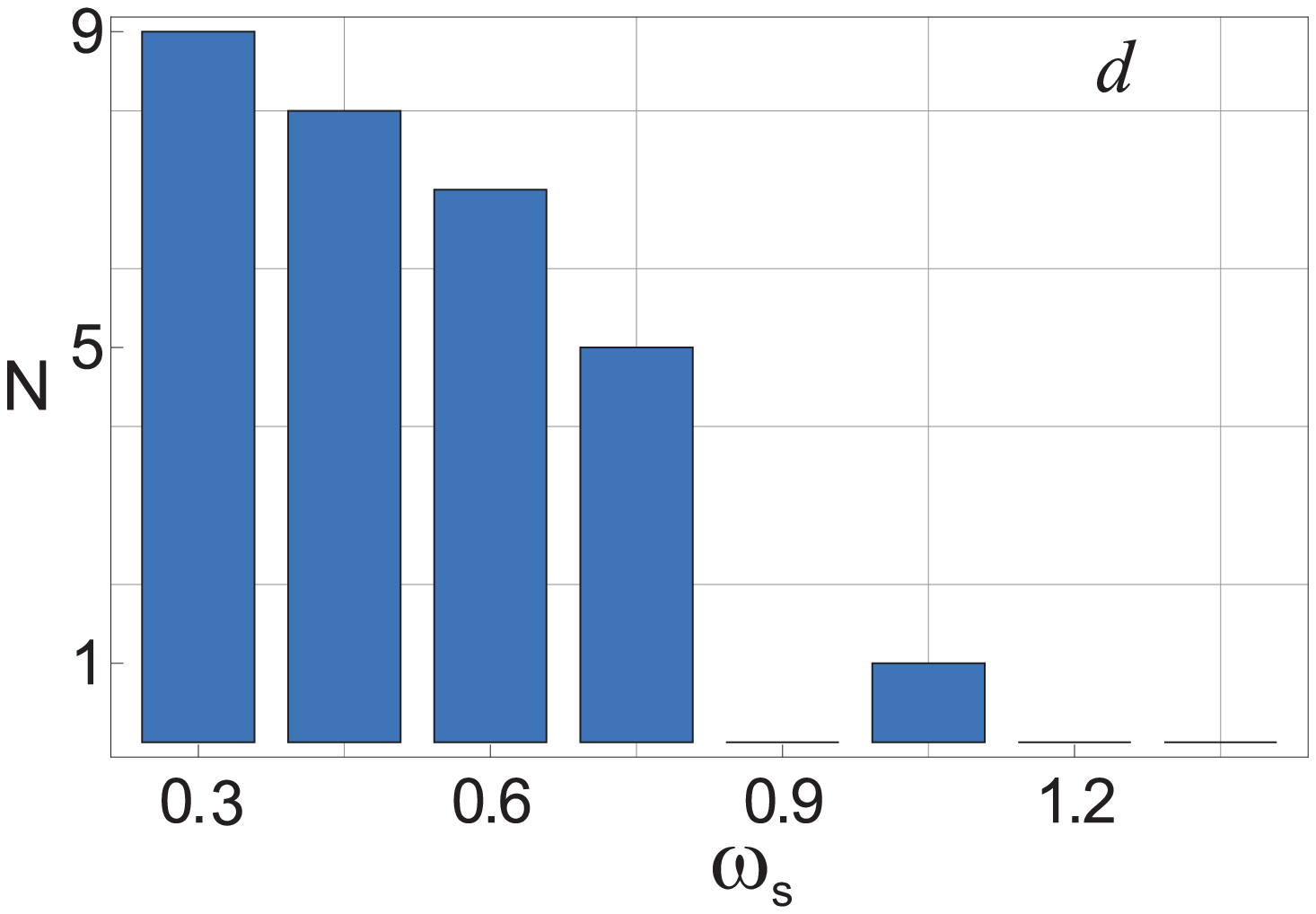}
\protect\caption{Histograms of basic GRB characteristics for the sample of GRBs presented in Table \ref{tab1}. On the Figures (a) to (d) we place distributions of (a) the Full Width at Half Maximum of the intensity of light curve pulse (FWHM), (b) the time of peak intensity for the observed pulse ($t_{\rm peak}$), (c) the maximal intensity ($J_m$) of the pulse and (d) the asymmetry indicator ($w$)}.
\label{fig01}
\end{figure*}

Fig. \ref{fig01} shows statistics of the measured parameters given in Table \ref{tab1}. As it can be seen in Fig. \ref{fig01} the values of observed parameters do not follow the Gaussian distribution. Due to the relatively small number of GRBs in the sample, here we are not able to discuss the power law indices of parameters (for detailed studies of these GRB properties see ~[\refcite{Norris96}], [\refcite{Hakkila08}]).

\subsection{Fitting procedure}

\begin{figure*}
\centering
\includegraphics[width=6cm]{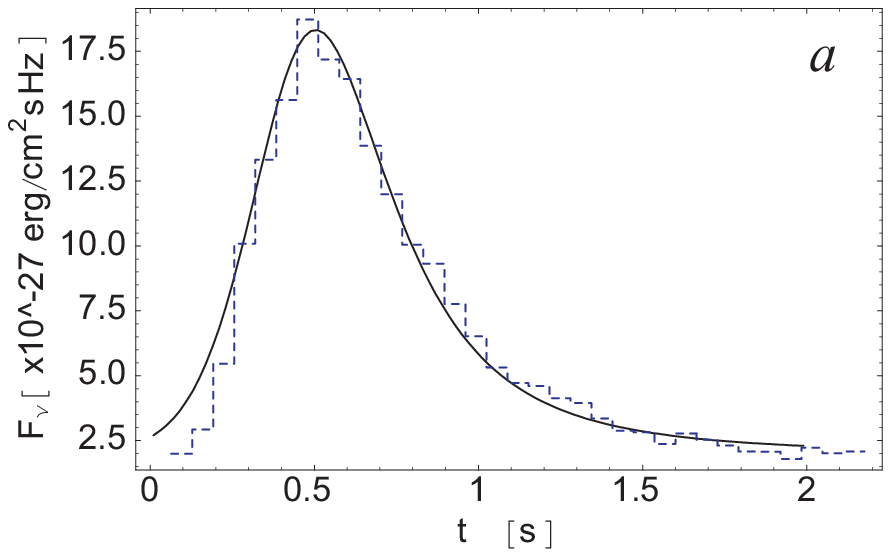}
\includegraphics[width=6cm]{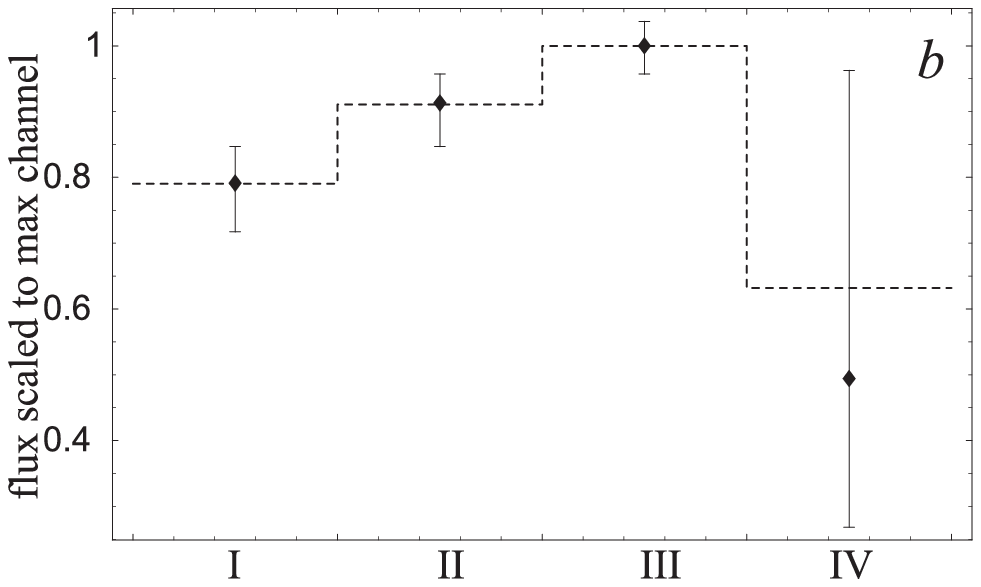}
\includegraphics[width=6cm]{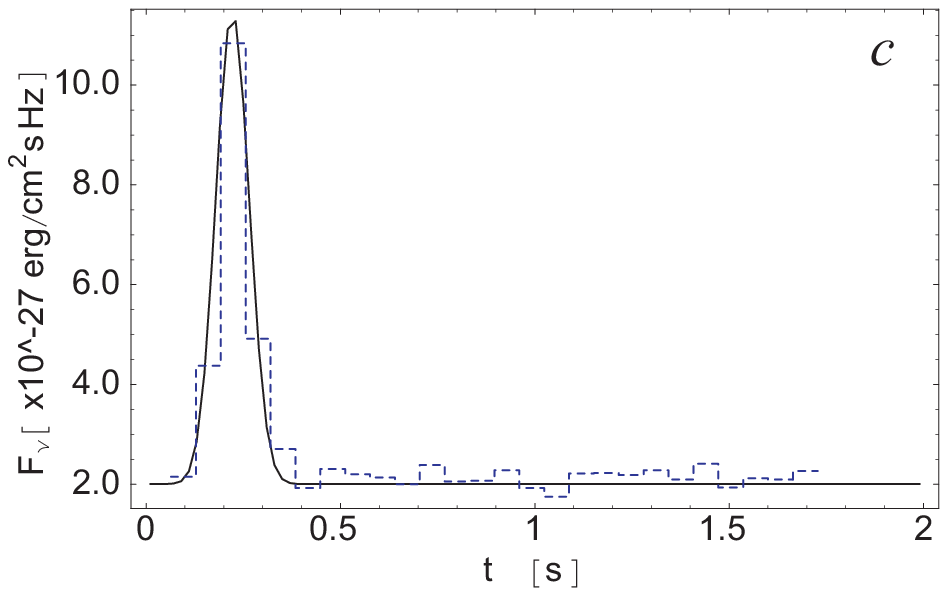}
\includegraphics[width=6cm]{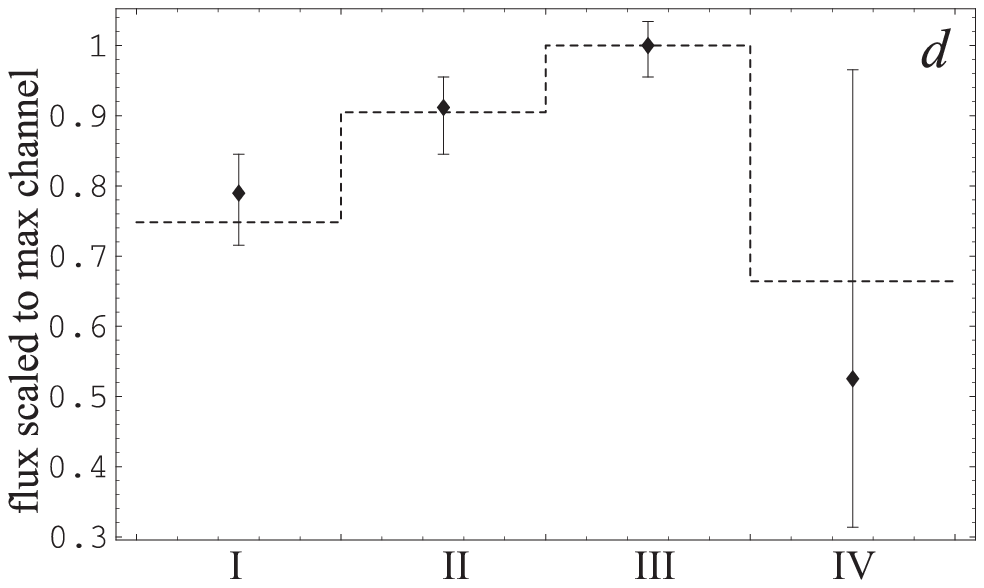}
\includegraphics[width=6cm]{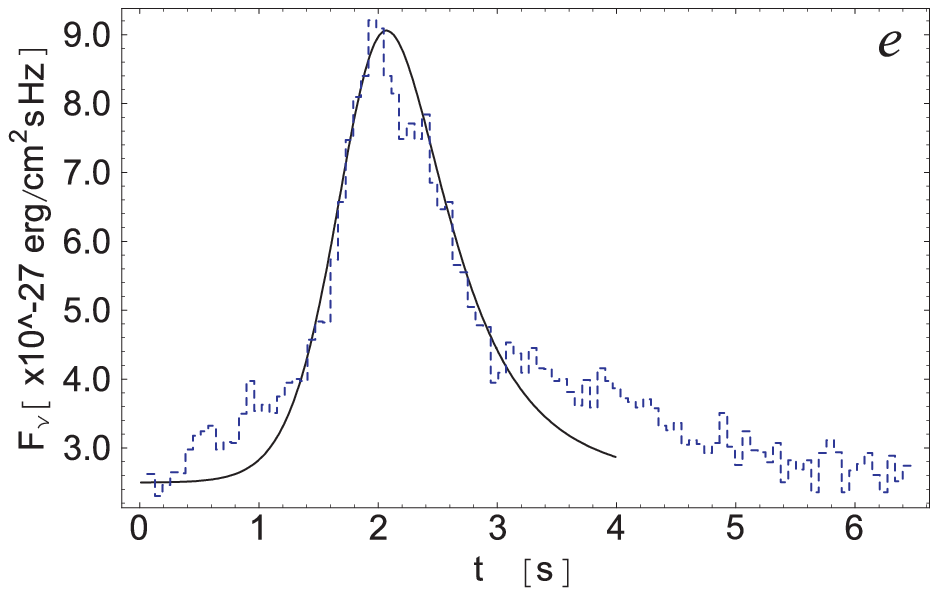}
\includegraphics[width=6cm]{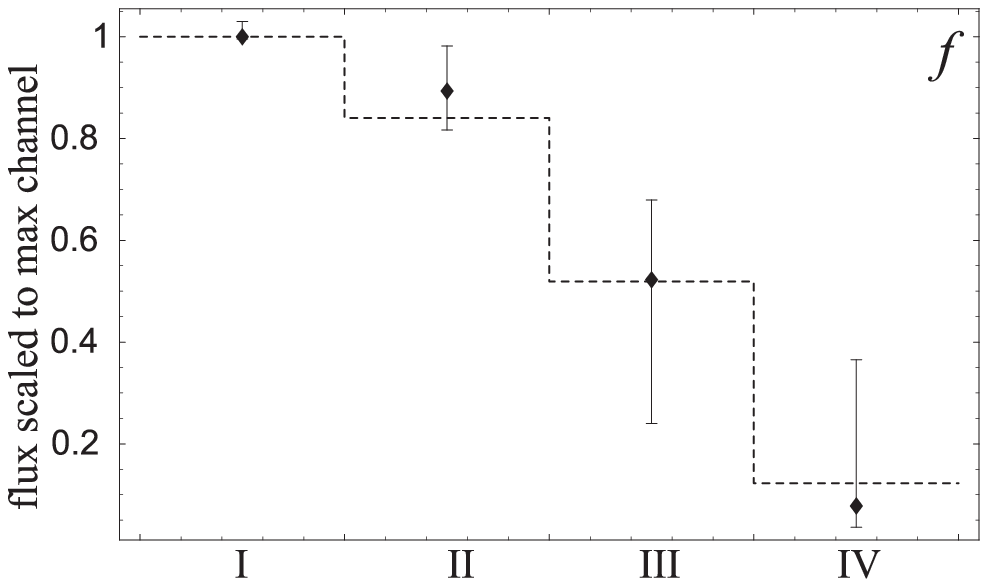}
\protect\caption{The light (dashed line - left panels) curves and the averaged spectral energy distribution in four BATSE channels I ch:(20-50)keV, II ch:(50-100)keV, III ch:(100-300)keV, IV ch:$>$300keV (full circles - right panels) for GRB911104, GRB911117 and GRB000508 (from top to bottom, respectively) fitted with the model (solid lines - left panels and dashed lines right panels).}
\label{fig02}
\end{figure*}

In order to follow the relativistic shell evolution and collision, we consider the case of only one shell expanding from the GRB core. It propagates through the surrounding media, which can contain a barrier with the mentioned Gaussian profile. We fit with our model light and energy curves of GRBs given in Table \ref{tab1}. In order to find the best fitting we vary the parameters of the faster shock and barrier. We consider the following parameters as free: the Lorentz factor $\Gamma_{0}$, the total initial ejected mass of the shell $M_{ej}$, the density of surrounding media $n_0$, the opening angle of the jet $\theta_m$, the distance of collision $R_c$ and the parameters $a$ and $b$ which describe the shape (height and width) of the density barrier. Additionally, we assumed that the barrier can move, thus we also put as a free parameter the Lorentz factor of the barrier $\Gamma_{b}$. In Fig. \ref{fig02} (left panels) we show  the best fit of three isolated pulses with different
shapes: GRB000508, GRB911104 and GRB911117. The light curves of these GRBs do not have a standard form,
i.e. the shape of pulses does not always follow the FRED (Fast Rise Exponential Decay) behavior.  As one can see from Fig. \ref{fig02} the shapes of the light and energy curves can be very well fitted with the model.
In right panels we show the best fit of the averaged spectral energy distribution (ASED)  taken from all four BATSE channels. Although the data for measured counts in the energy channels are with large uncertainties, the fit of ASED can be used for the confirmation of the validity of the GRB light curve fit, since the same parameters are used to fit both curves.

The coefficients $\xi, \xi_{e}$ and $\xi_{b}$\footnote{$\xi$ describes fraction of total shell
energy that has been converted into radiation, $\xi_{e}$ fraction of the total shell energy devoted to the electron plasma component and $\xi_{b}$ fraction of total shell energy contained in the magnetic field.}, determine the radiation efficiency for expanding relativistic shell. The synchrotron radiation is directly proportional to intensity of the magnetic field, as well as the energy of the electron component of plasma. Consequently, if one puts high values for those three components it will increase the intensity of radiation and also the intensity of light curve pulses. However, the conversion from the kinetic to radiative energy is the process with relatively low efficiency (see e.g. [\refcite{Eichler05}]), therefore the above coefficient must not exceed 10 to 20\% [\refcite{Piran05}]. They are expected to be smaller than 0.2. In order to reduce the number of free parameters in the fitting procedure we fixed them to $\xi = 0.1$, $\xi_e = 0.2$, $\xi_b = 0.2$ (see also [\refcite{Zhang07}]). Moreover, the different values of these parameters will mainly affect the intensity of GRB light curve, but not the shape which is mainly considered in the fitting procedure.

Also, the distribution of electrons in the shell follows a power law function where the index of the electron
distribution $p$ usually takes a value between 2 and 3 [\refcite{Gallant99}]. In the fitting we fix it to $p = 2.5$
which well reproduce the obtained values from the fits of the GRB light and energy curves.
In the fitting procedure we used $\chi^2$ minimization.

\section{Results and discussion}

\begin{figure*}
\centering
\includegraphics[width=4cm]{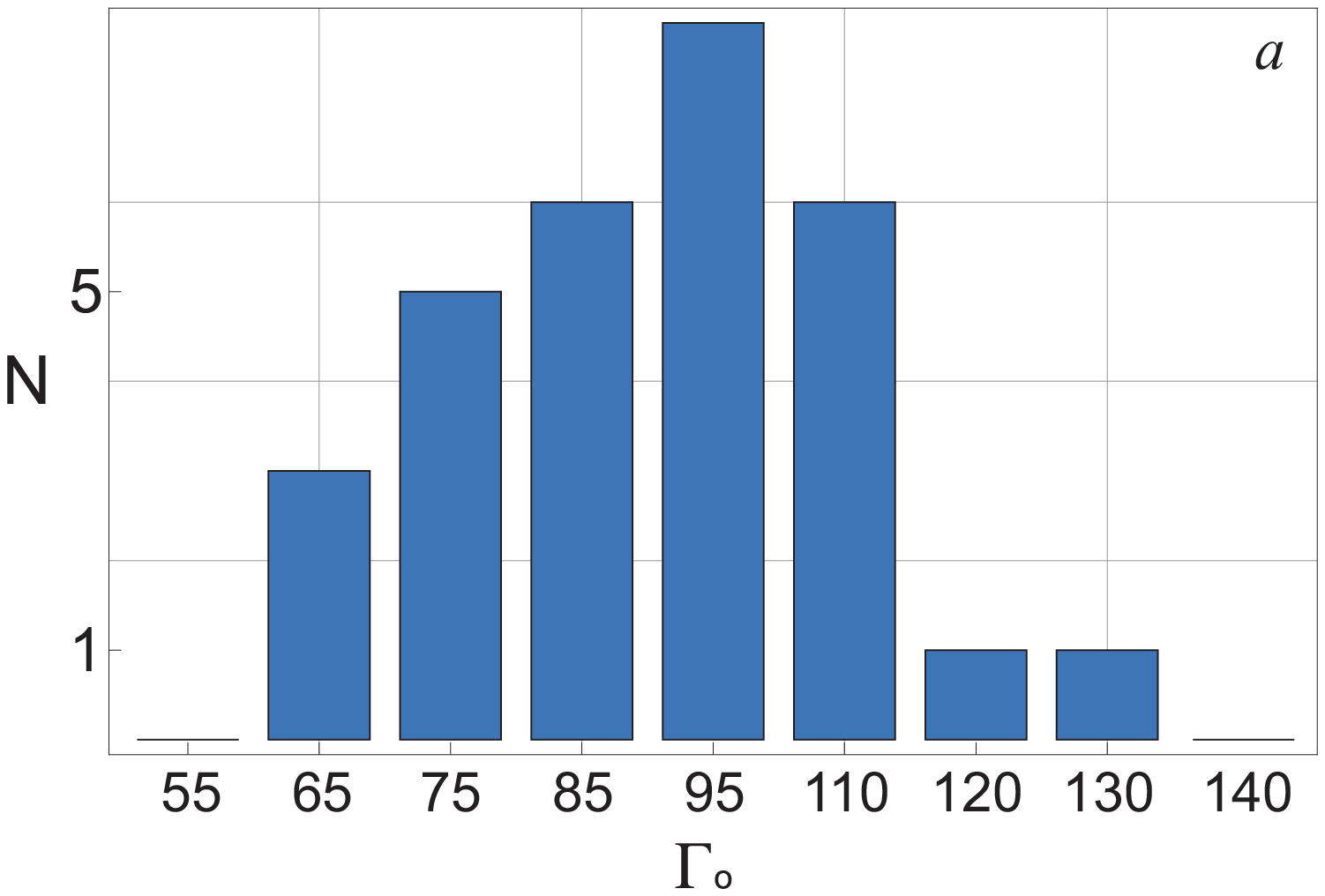}
\includegraphics[width=4cm]{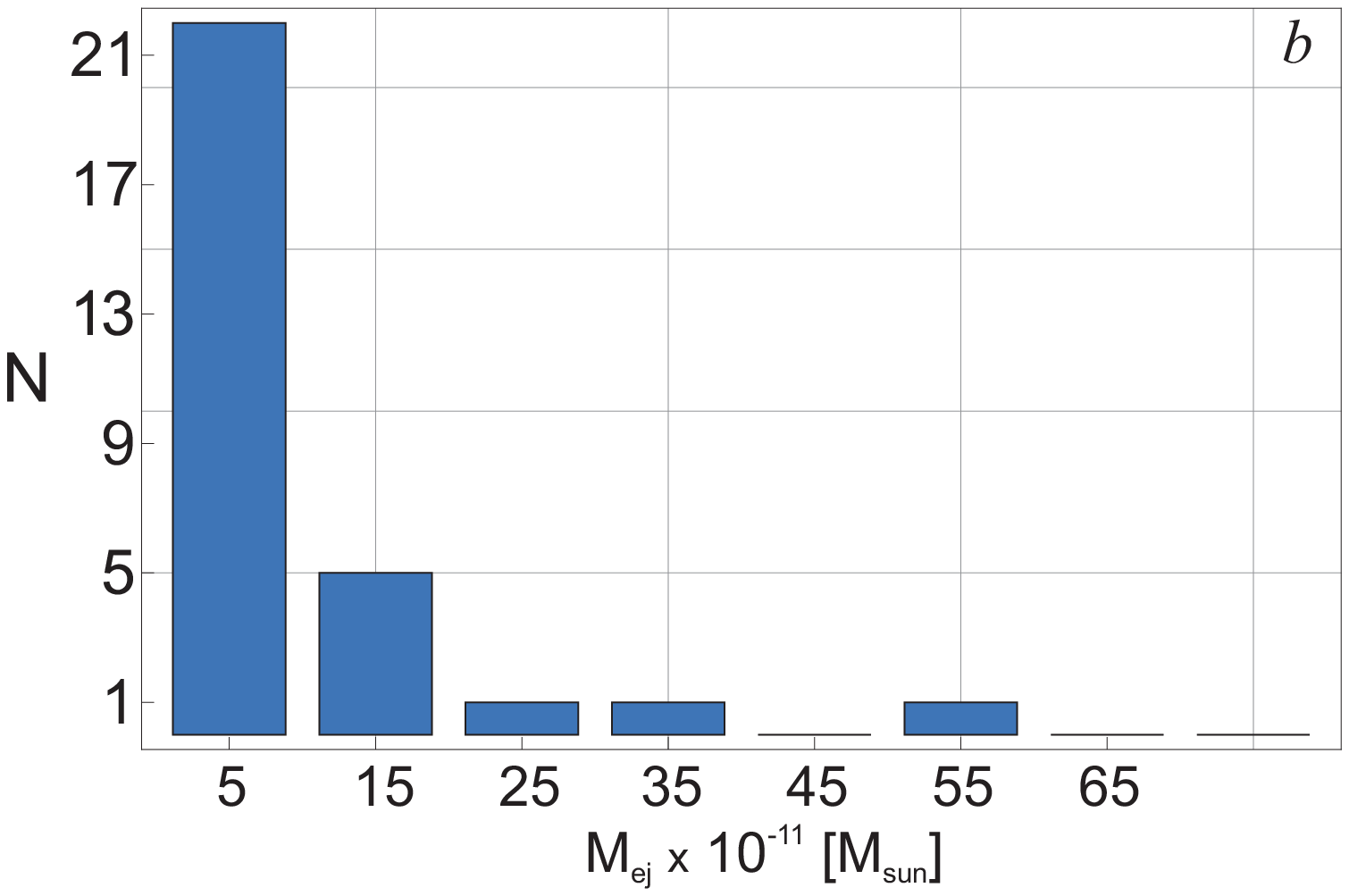}
\includegraphics[width=4cm]{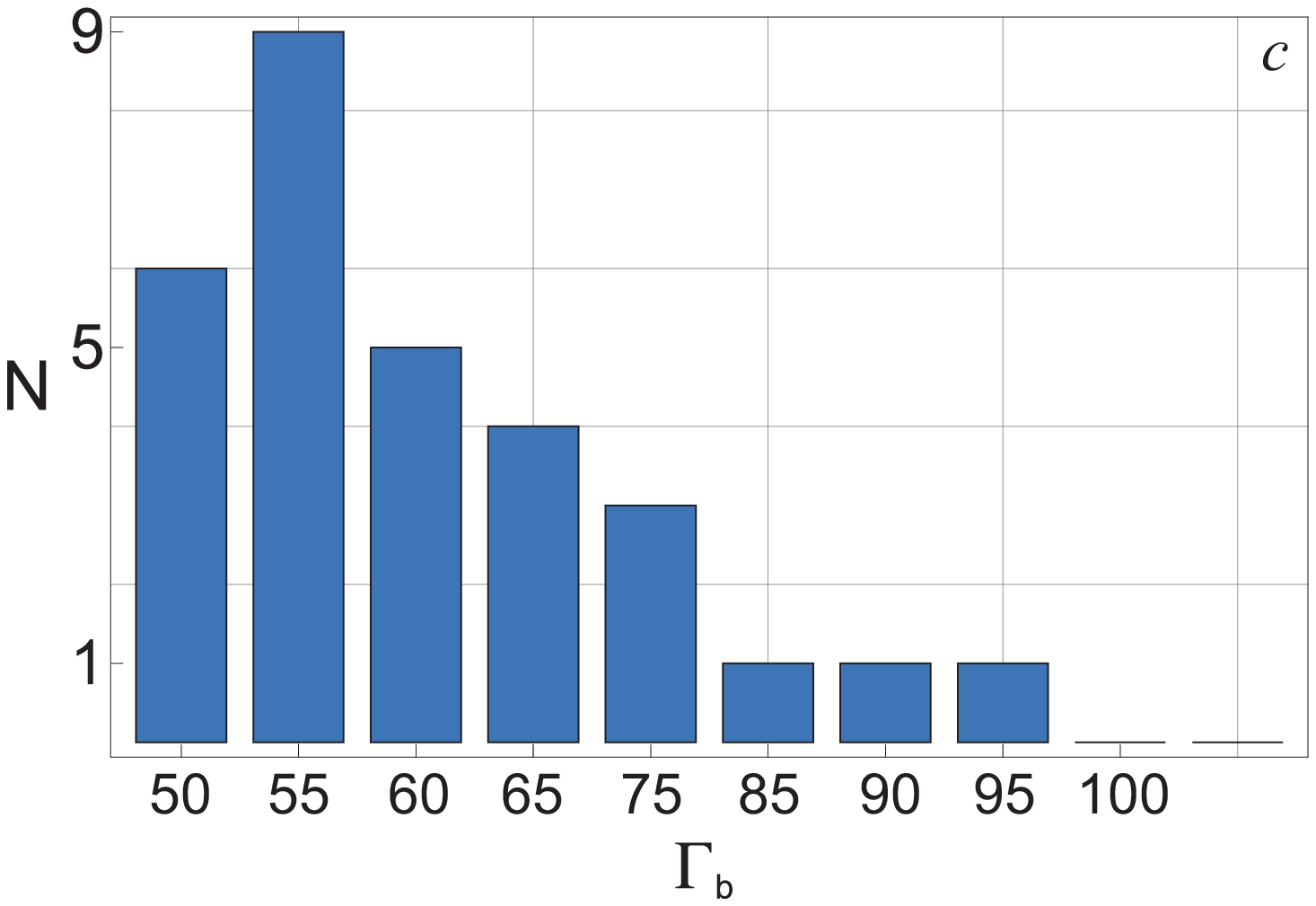}
\includegraphics[width=4cm]{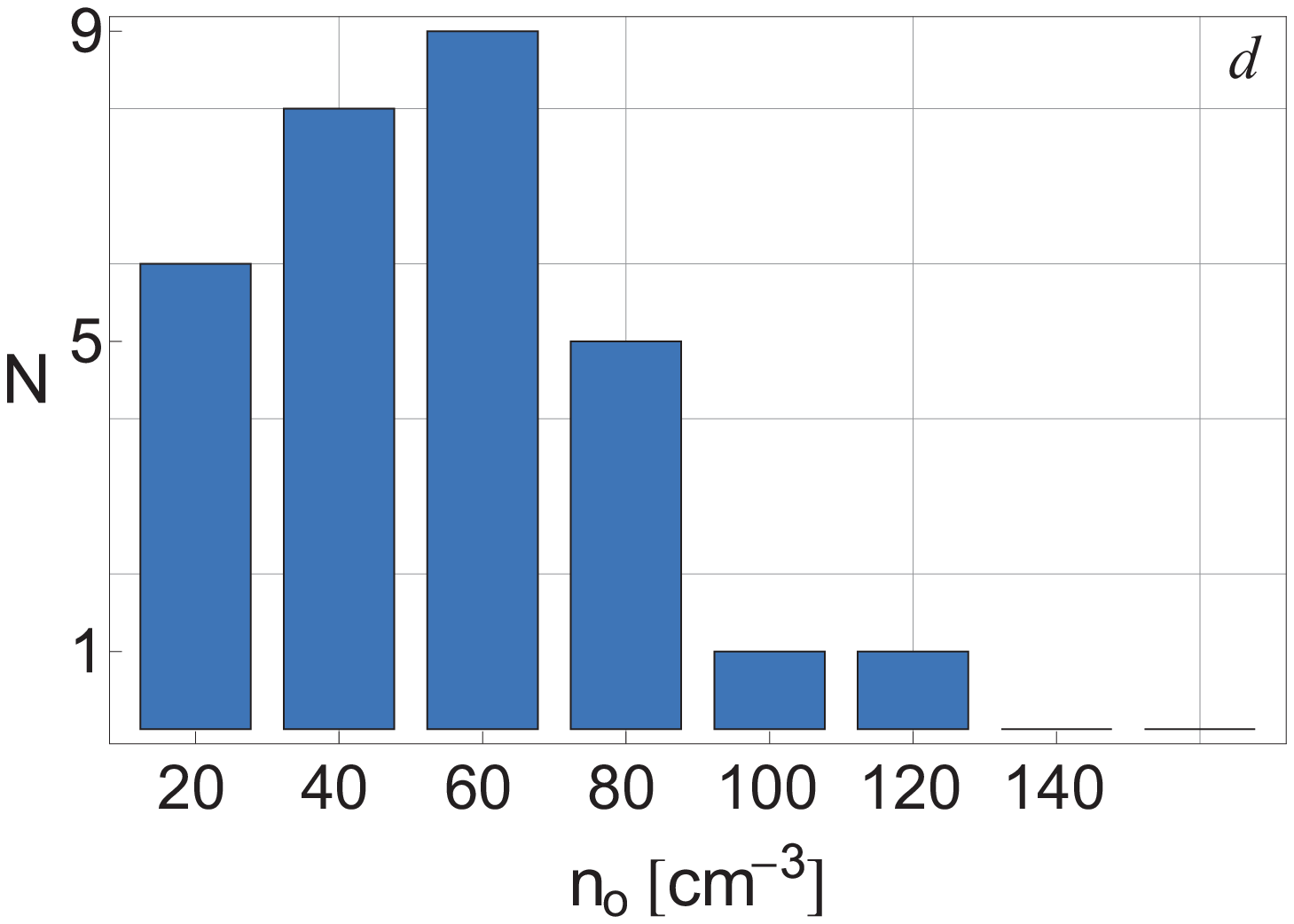}
\includegraphics[width=4cm]{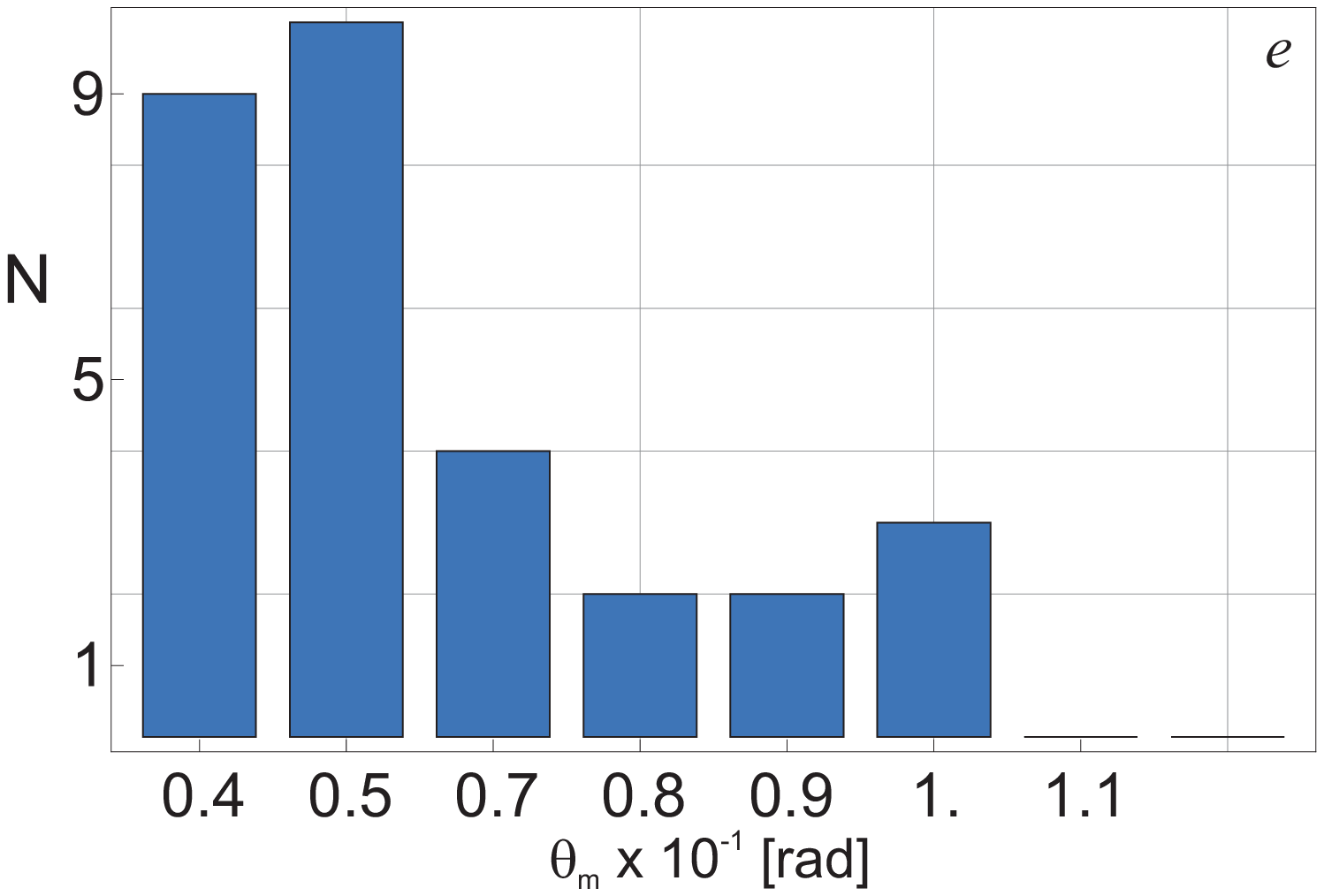}
\includegraphics[width=4cm]{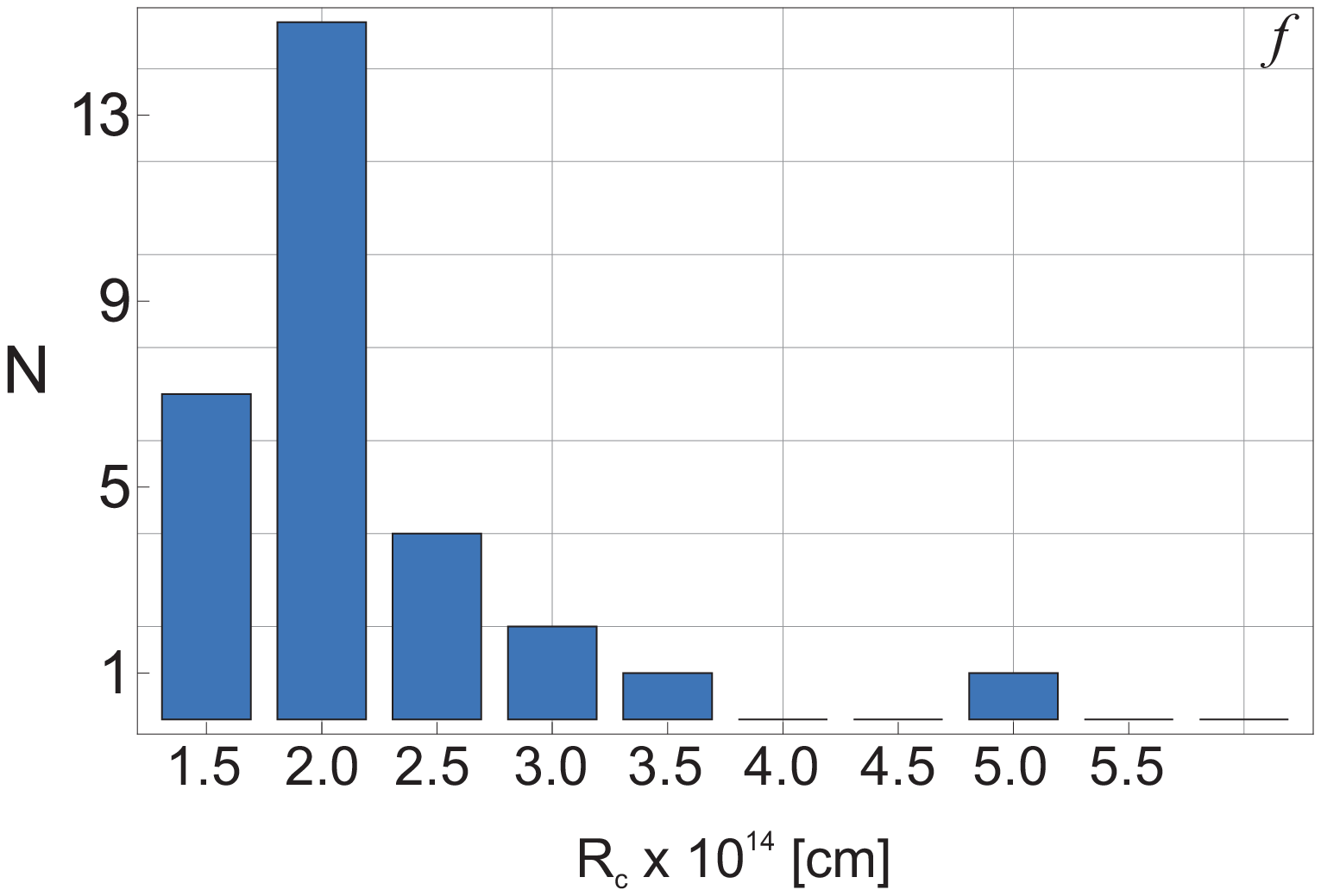}
\includegraphics[width=4cm]{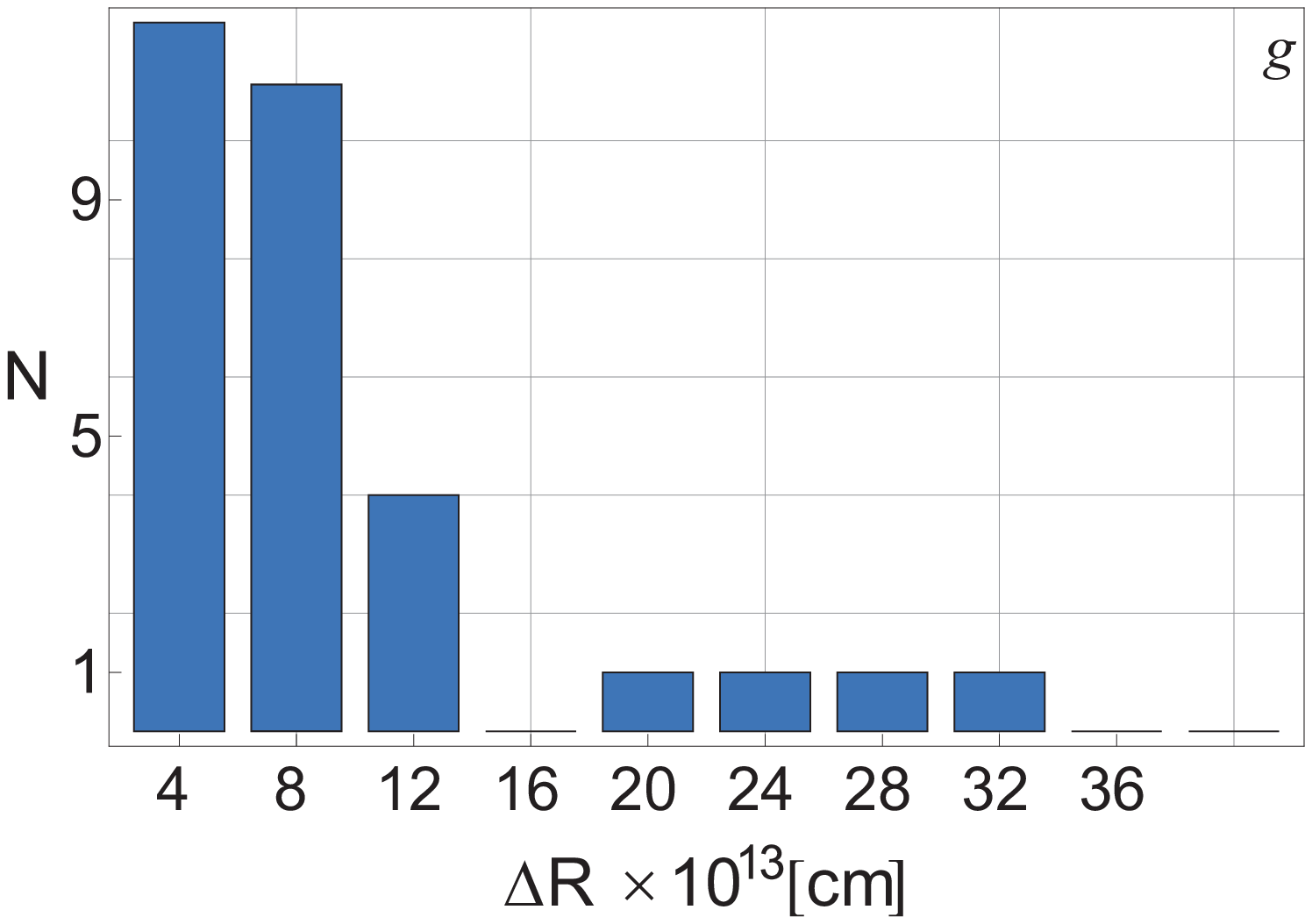}
\caption{Histograms for obtained parameters from the best fit:  $\Gamma_0$, $M_{\rm ej}$, $\Gamma_b$, $n_{o}$, $\theta_{m}$, $R_{c}$ and $\Delta R$ (a-g panels, respectively).}
\label{fig03}
\end{figure*}

The parameters of the shells and barriers for 30 GRBs obtained from the best fittings are given in Table \ref{tab2}. The distribution of parameters are presented in Fig. \ref{fig03}. As it can be seen from Table \ref{tab2} (also see Fig. \ref{fig03}) the most of GRB light curve pulses in our sample indicate an initial Lorentz factor around 95, with the mean value of about 93. Cases with $\Gamma_{0}<90$ could be explained if we consider the physical mechanism of the collisions. In the first place, the initial value of the Lorentz factor directly determines the initial energy and velocity of the incoming shell. So, for those shells (higher $\Gamma_0$) produced light curve pulses will be more intense and short lasting. On the contrary, smaller $\Gamma_{0}$ will produce low intensity and long lasting pulses. This is in a good agreement with the observations of the pulse width - luminosity correlation found in [\refcite{Hakkila08}].

\begin{table}
\tbl{The internal shell and barrier parameters obtained from the best fitting of light and energy curves for the sample of 30 GRBs. In the last two rows we put the mean and appropriate standard deviation of parameters.}
{\begin{tabular}{@{}ccccccccc@{}} \toprule
parameter & $\Gamma_{0}$ & $M_{\rm ej}$ & $\Gamma_{b}$ & $n_{0}$ & $\theta_{m}$ & $R_{c}$ & $\Delta R$ & $n_{b}$ \\
units & - & $\cdot 10^{-11}[M_{\rm sun}]$ & - & $[\rm cm^{-3}]$ & $[\rm rad]$ & $ \cdot 10^{14}\rm [cm]$ & $ \cdot 10^{13}\rm [cm]$ & $\cdot 10^{4} \rm [cm^{-3}]$ \\ \colrule
GRB910629 & 109 & 5.5 & 50 & 43 & 0.05 & 2.0 & 5.1 & 43. \\
GRB911104 & 111 & 10. & 43 & 63 & 0.06 & 2.5 & 5.1 & 56.7 \\
GRB920715 & 103 & 1.3 & 51 & 33 & 0.06 & 2.2 & 5.8 & 33. \\
GRB920720 & 90 & 25. & 49 & 70 & 0.06 & 1.5 & 5.8 & 105. \\
GRB920808 & 110 & 8. & 50 & 33 & 0.07 & 1.9 & 7.8 & 8.3 \\
GRB920811 & 111 & 1.1 & 65 & 65 & 0.06 & 1.2 & 2.2 & 58.5 \\
GRB920830 & 71 & 13.5 & 45 & 37 & 0.1 & 3.5 & 25. & 1.1 \\
GRB920912 & 100 & 10. & 74 & 70 & 0.05 & 2.0 & 3.5 & 4.2 \\
GRB920924 & 125 & 2. & 50 & 50 & 0.04 & 2.5 & 3.8 & 150. \\
GRB921021 & 95 & 5. & 65 & 20 & 0.08 & 3.5 & 21.7 & 0.6 \\
GRB921207 & 73 & 53. & 63 & 75 & 0.1 & 2.5 & 10. & 150. \\
GRB921208 & 71 & 3.8 & 49 & 70 & 0.07 & 1.9 & 7.3 & 4.9 \\
GRB921222 & 87 & 2.5 & 70 & 20 & 0.06 & 2.1 & 7.8 & 6. \\
GRB950129B & 84 & 4. & 70 & 20 & 0.05 & 1.8 & 7.5 & 100. \\
GRB950211B & 73 & 18. & 56 & 90 & 0.05 & 1.7 & 8.3 & 4.5 \\
GRB960111 & 95 & 10. & 68 & 45 & 0.08 & 1.9 & 7.7 & 1.8 \\
GRB960207 & 109 & 6.8 & 63 & 57 & 0.06 & 1.5 & 3.2 & 39.9 \\
GRB960229 & 98 & 5.2 & 61 & 75 & 0.057 & 1.7 & 4.8 & 9.8 \\
GRB960311 & 110 & 13. & 77 & 19 & 0.04 & 1.9 & 5.2 & 9.5 \\
GRB960409 & 75 & 18. & 52 & 110 & 0.1 & 5.5 & 30. & 5.5 \\
GRB960418 & 75 & 3. & 48 & 50 & 0.09 & 2.7 & 10.7 & 50. \\
GRB960524 & 87 & 2.7 & 65 & 50 & 0.09 & 2.1 & 8.8 & 335. \\
GRB960528 & 91 & 5.5 & 56 & 43 & 0.06 & 2.0 & 13.3 & 2.2 \\
GRB960530 & 62 & 40. & 46 & 31 & 0.055 & 3.8 & 30.1 & 0.6 \\
GRB960613 & 87 & 13. & 63 & 65 & 0.04 & 2.4 & 10.2 & 3.9 \\
GRB960617B & 103 & 3. & 87 & 50 & 0.07 & 2.3 & 8.3 & 150. \\
GRB970424 & 99 & 0.9 & 53 & 35 & 0.061 & 1.5 & 3. & 105. \\
GRB991105 & 99 & 6. & 57 & 69 & 0.07 & 2.3 & 6.8 & 4.8 \\
GRB991213 & 75 & 5. & 55 & 10 & 0.05 & 2.1 & 9.3 & 9. \\
GRB000107 & 115 & 4.3 & 90 & 10 & 0.05 & 1.5 & 3.3 & 10. \\
\hline
Mean & 93 & 10 & 60 & 50 & 0.064 & 2.26 & 9.4 & 48 \\
Deviation & 13.5 & 7.6 & 9.6 & 19.4 & 0.014 & 0.57 & 5.0 & 52 \\
\botrule
\end{tabular} \label{tab2}}
\end{table}

The total released isotropic energy of a GRB lies in the range from $8 \times 10^{47}$ erg (GRB 980425
associated with supernova SN1998bw, see e.g. Ref~[\refcite{Galama98}], [\refcite{Pian00}]), to $2 \times 10^{54}$ erg (GRB 990123 Ref.~[\refcite{Andersen99}]). Note here that using the conical jet model [\refcite{Frail01}] found that the gamma-ray energies are clustered in a narrow interval around $5\times 10^{50}$ erg. The mass of the ejected shell in the most GRB light curve pulses is in the interval of $ \sim 10^{-11}-10^{-10} M_{\rm sun}$, with a mean value of $10^{-10} M_{\rm sun}$. These could be used to calculate the initial energy of a particular shell in the moment of the ejection, giving values in the range from $10^{45}$ to $10^{47}$ergs. The sum of energies of all ejected shells during the GRB event can indicate the
total energy released by the central engine in the particular GRB.

Additionally we assume that the barrier is moving, with the Lorentz factor $\Gamma_b$. In the first approximation we take that its velocity is constant until the collision. This parameter has multiple influence mainly on the intensity and width of the created light curve pulse, as well as on the FRED pulse shape. For the faster barriers interaction is long-lasting, but with a low intensity. On the other hand, if the barrier is moving significantly slower than the incoming shell, it will cause a more intense interaction followed by the short-lasting and very strong pulses. The FRED shape is more dominant in the former case.

The density of the ISM in the region around the central engine is assumed to be homogeneous $(s = 0)$, with a density approximately an order of magnitude higher than at the distances where afterglow starts (see for example [\refcite{Kobayashi04}]). We obtained values from several to few tens of particles per cm$^{-3}$ (see Table\ref{tab2}) that is in the expected range $n_{0} \simeq (10^{0} - 10^{3})$ cm$^{-3}$. The distribution of $n_{0}$ (Fig. \ref{fig03}d) for our sample has a bell-like shape with the maximal value around 60 cm$^{-3}$. This values is more appropriate to hypernova then to a merger scenario, because in the merger scenario one can expect significantly smaller densities [\refcite{Hakkila08}].

Similar distribution is obtained for opening angle of relativistic shells $\theta_{m}$, which is considered to be constant during the evolution. The most probable values are around 0.05 radians ($\approx 3^{o}$)
(see [\refcite{Frail01}] and reference therein). Existence of this particular value is determined by physical processes in
the vicinity of the GRB central engine. Namely, if one take a higher value of $\theta_{m}$, the resulting pulse
is broader and a slow decay feature is much more visible than in the case of smaller angles, where
the  pulse is thinner and has a symmetrical shape. That is in a good agrement with engaged physical processes during the shell expansion.

We suppose that the shell interaction occurs mainly close to the GRB engine, (distance $R_c\sim10^{14}\ \rm cm$),
as it was proposed in the internal shock scenario (see [\refcite{Piran05}]). This parameter
has very small influence on the shape of GRB pulses, but has an influence on the intensity of pulses. Pulses produced in a collision closer to the GRB engine are more intense than ones originating at larger distances.

The parameters of the Gaussian of the density distribution which describe the structure of the barrier, the width at
the half maximum $b$ and the density in the central part $a$, both influence the shape of GRB light curve pulses.
They are translated into appropriate variables, number density $n_b$ and width $\Delta R$ of the barrier,
respectively, using the connections given by Eq. \ref{eq9} and \ref{eq10}:

\begin{equation}
n_{b} = n_0 (1+a)
\label{eq9}
\end{equation}

\begin{equation}
\Delta R = 2 b \sqrt{\ln \frac{2 a} {a - 1}}
\label{eq10}
\end{equation}

Note here that in the case of dense barrier one can expect $n_{b}/n_0>>1$ (see also Table \ref{tab2}), consequently $a>>1$ and Eq. \ref{eq10} can be rewritten as $\Delta R = 2 b \sqrt{\ln {2}}$. The influence of the barrier parameters on pulse profiles is following: a barrier with narrow width and high number density will have a strong interaction with fast shell, producing symmetrical and intense light curve pulses. On the contrary, a barrier with larger width  and lower number density will cause small intensity pulses with high asymmetries.

In general, comparing the obtained values of parameters (Table \ref{tab2}) for different GRBs, one can conclude
that there is no significant difference between them even when the shapes and durations of GRB pulses are different. This
suggests that the nature of GRBs is similar and that there should be no big difference between the physical
conditions of GRB progenitors (see [\refcite{Ghirlanda11}]). On the other hand, the barrier density distribution can differ from the Gaussian, that is assumed here, and it may reflect the values of basic parameters. But in any case, one can expect that the density distribution of the barrier has to be taken into account in the shock model.

\subsection{Connection between the shell parameters}

\begin{figure*}
\centering
\includegraphics[width=6cm]{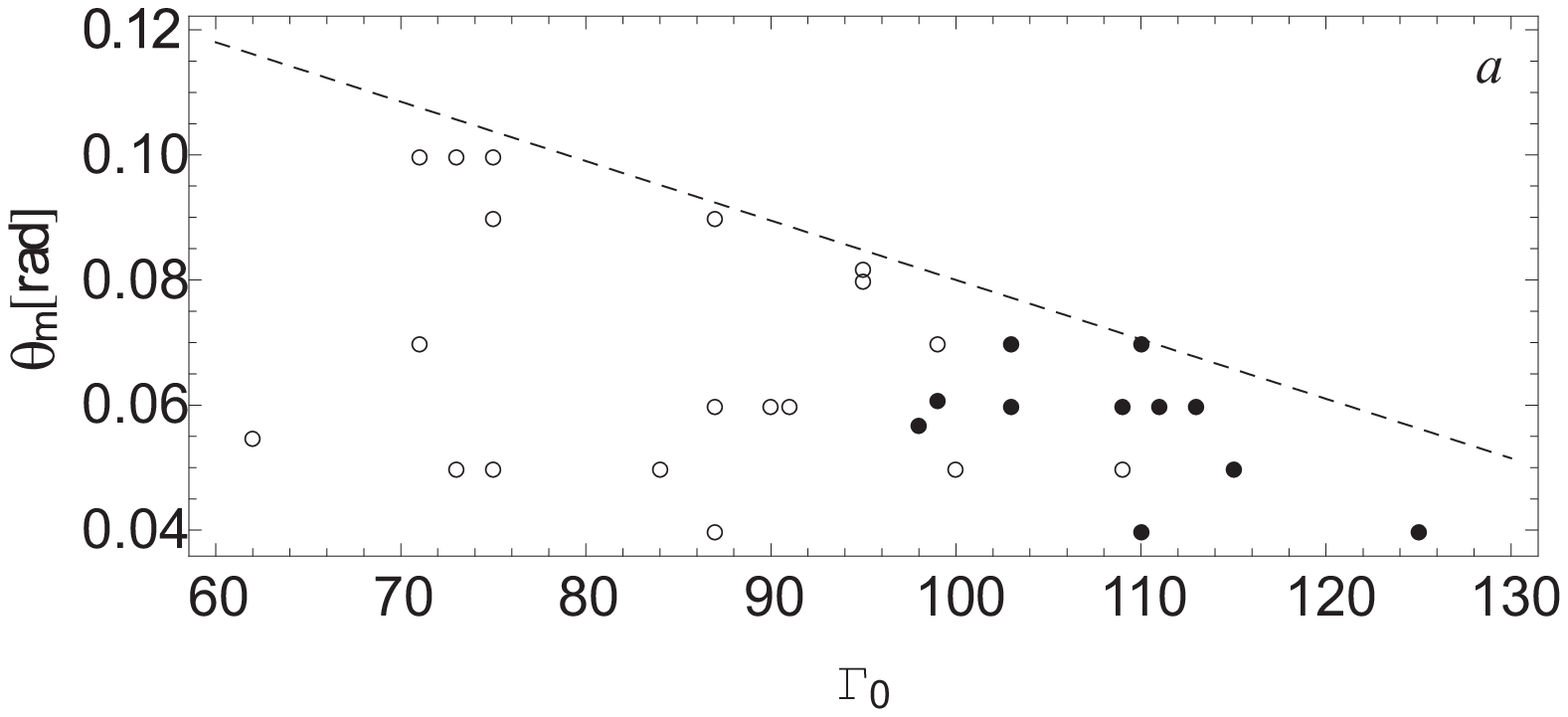}
\includegraphics[width=6cm]{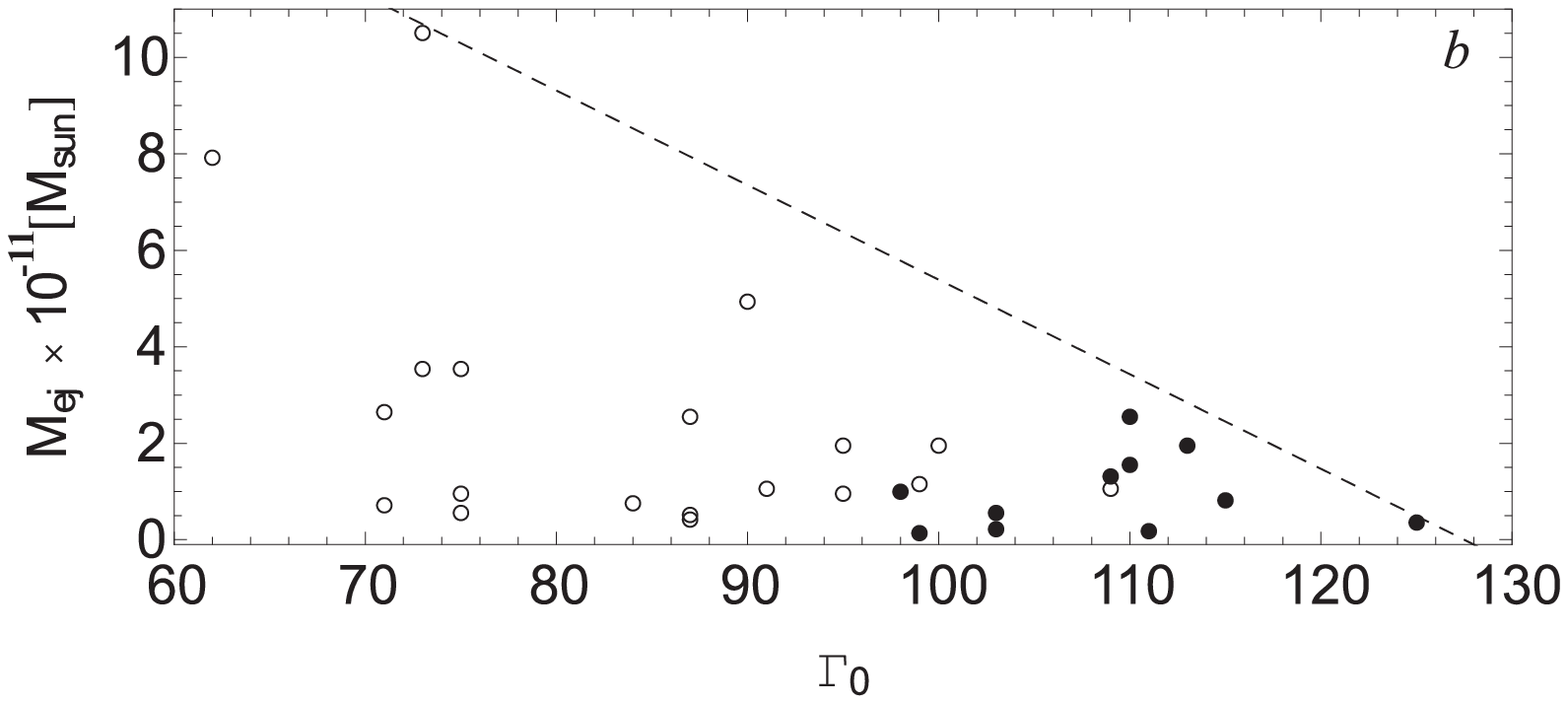}
\caption{Parameter dependance: $\theta_{m}$ vs. $\Gamma_0$ (a) and $M_{\rm ej}$ vs. $\Gamma_{0}$ (b). The long (t$>2$s) GRBs are denoted with full circle and short with open circle. Dashed line presents the border above which there is no parameter values for this sample of GRBs.}
\label{fig04}
\end{figure*}

\begin{figure*}
\centering
\includegraphics[width=6cm]{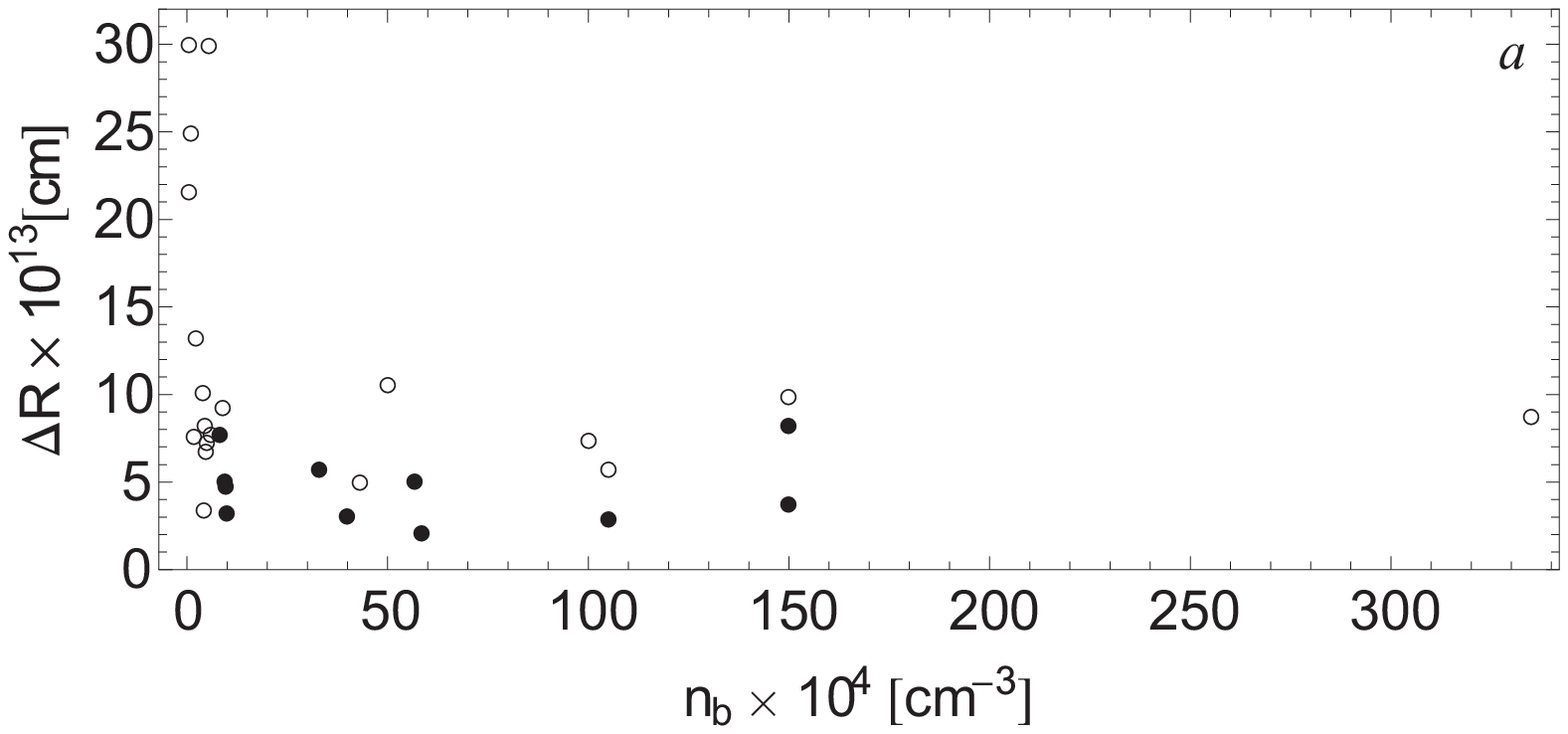}
\includegraphics[width=6cm]{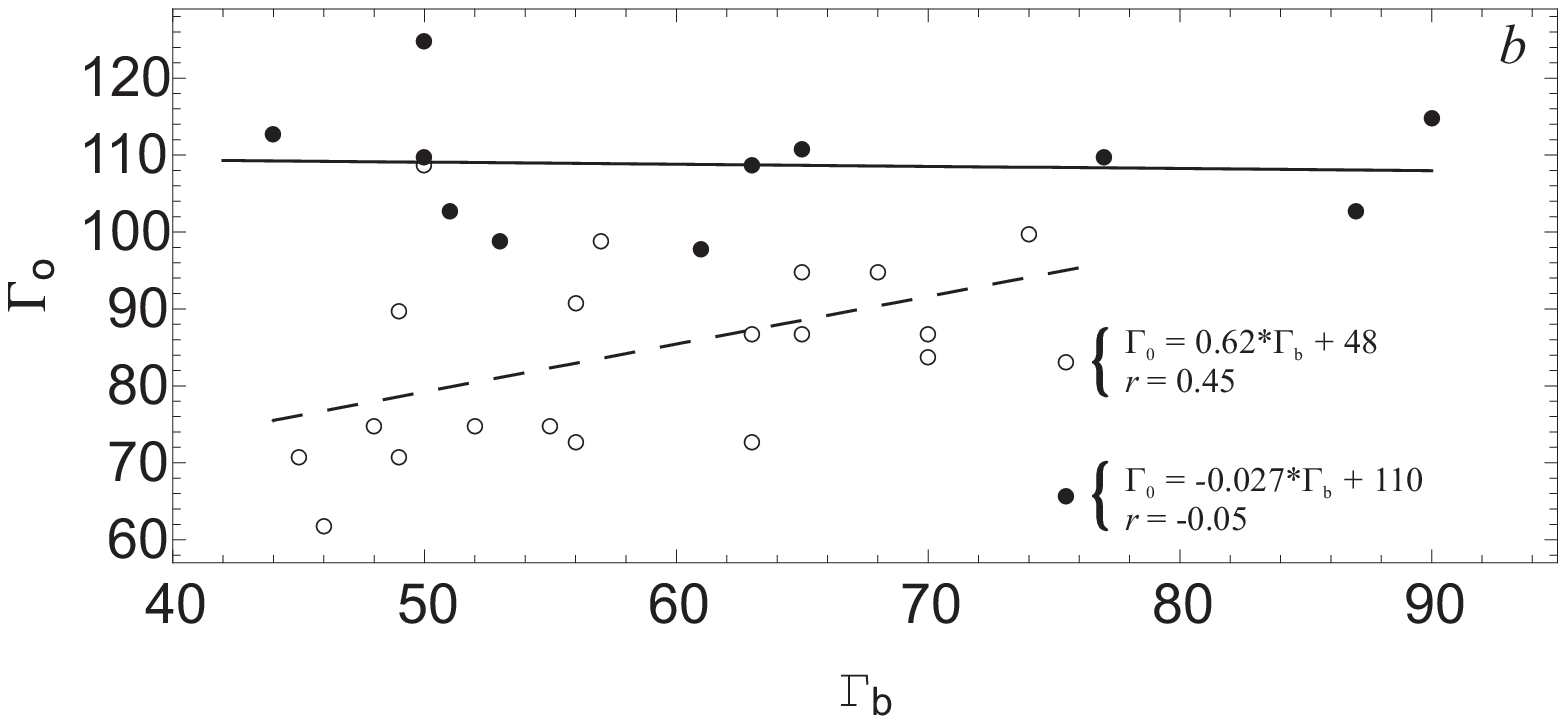}
\caption{Parameter dependance: $\Delta R$ vs. $n_{b}$ (a) and $\Gamma_{0}$ vs. $\Gamma_{b}$ (b), with $r$ designating the correlation coefficient. The notation for short and long pulses is same as in Fig. 4.}
\label{fig05}
\end{figure*}

In order to find physical meaning of the obtained parameters, we explore correlations between them.
In Table \ref{tab2} the parameters are divided in two groups: one that describes a shell and other connected with the barrier. We expect that parameters from those two groups are not in  correlation, since they are independent. The results are presented in Figs. \ref{fig04} and \ref{fig05}, where  we separately denoted the long (t$>$2s) and short (t$<$2s) GRBs with open and full circles, respectively.
As one can see in Figs. \ref{fig04} and \ref{fig05} there are no strong correlations between these quantities, but only in some cases, there is a slight connections between different parameters, as e.g. $\theta_{m}$ vs. $\Gamma_{0}$ (Fig.\ref{fig04}a),
$M_{\rm ej}$ vs. $\Gamma_0$ (Fig. \ref{fig04}b). In the case of the barrier parameters a slight correlation can be found in
$\Delta R$ vs. $n_b$ (Fig. \ref{fig05}a). Additionally, we examine the correlation which may be established between
Lorentz factors of the incoming fast shock wave $\Gamma_0$ and the moving barrier $\Gamma_b$, presented in Figure
\ref{fig05}b.

Weak correlations or non-correlations can be noticed in each case particulary. For example, with larger
opening angle of the shock wave $\theta_m$, the volume of the shell increases, that causes smaller initial Lorentz factors of the shell and {\it vice versa} (see Fig. \ref{fig04}a). For the certain shell energy, the increase of the space angle $\theta_m$ causes the increase of the ejected mass $M_{\rm ej}$, and that causes the decrease of the initial $\Gamma_{0}$. This is in a agrement with the trends presented in Figure \ref{fig04}b, where one can see that for a higher value of $M_{\rm ej}$, the initial Lorentz factor tends to be smaller. Of course, one can not expect to see obvious confirmation for this conclusion with such small sample of examined GRBs, but rather signs of trends for mentioned dependence.

In the case of the barrier parameters, there is an indication of the connection
between the particle number density $n_b$ and width of barrier $\Delta R$ as it is shown in Fig. \ref{fig05}a, as $\Delta R\sim 1/n_b$. Physically one can expect such situation, i.e. for a broader barrier, the density of barrier tends to be smaller and {\it vice versa}.

In Fig. \ref{fig05}b the $\Gamma_0$ vs. $\Gamma_b$ is presented. There is no global correlation between these Lorentz factors, but taking into account only short pulses there is some indication that for a faster barrier the $\Gamma_0$ of the shell is higher. The dashed and solid line in Fig. \ref{fig05}b present the linear dependance of the given parameters, indicating that the incoming shell must have higher $\Gamma_0$ than the slower moving barrier, providing just enough necessary conditions for event of collision to happen.

\subsection{Connection between shell parameters and observed pulse parameters}

\begin{figure*}
\centering
\includegraphics[width=6cm]{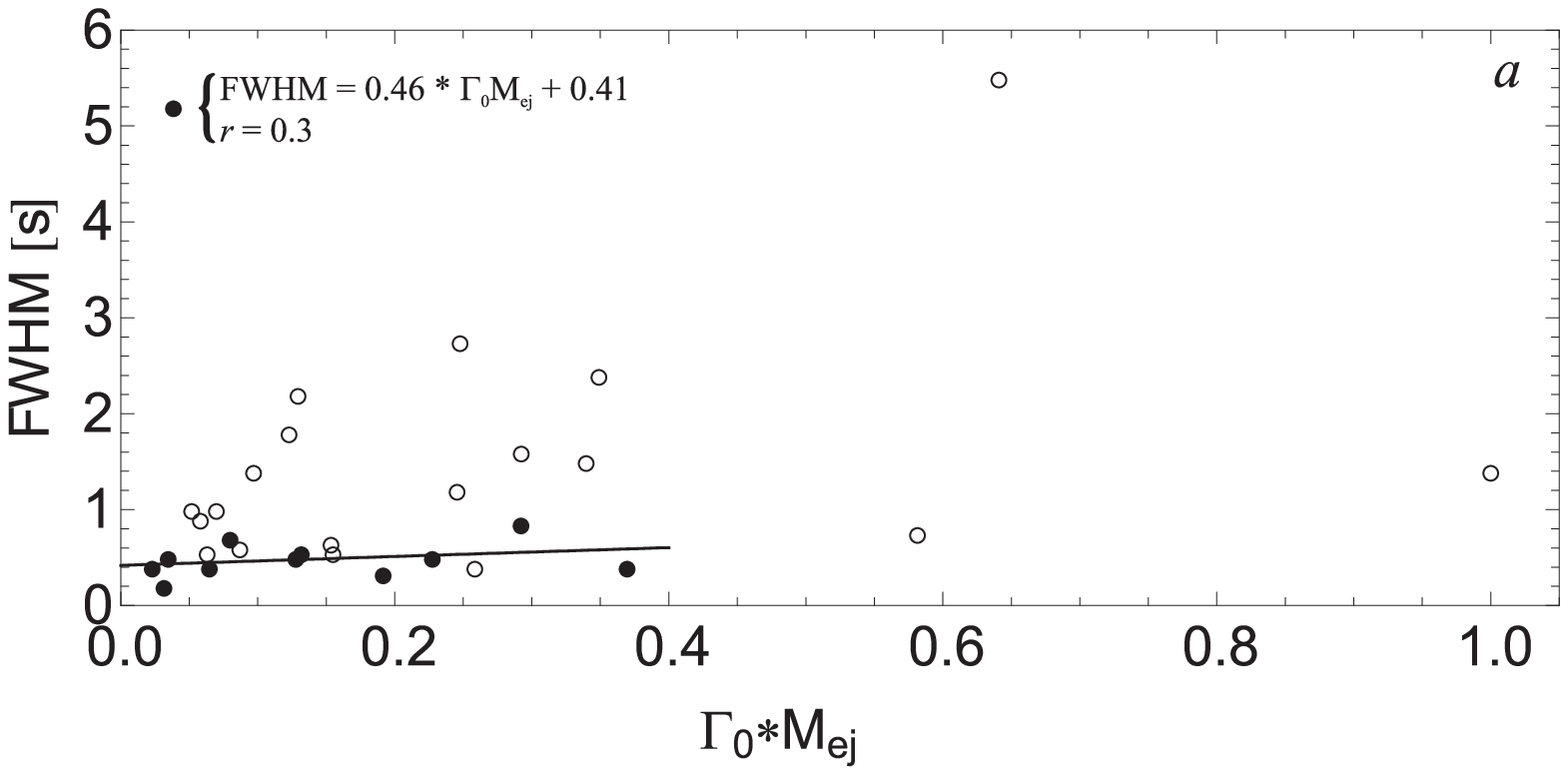}
\includegraphics[width=6cm]{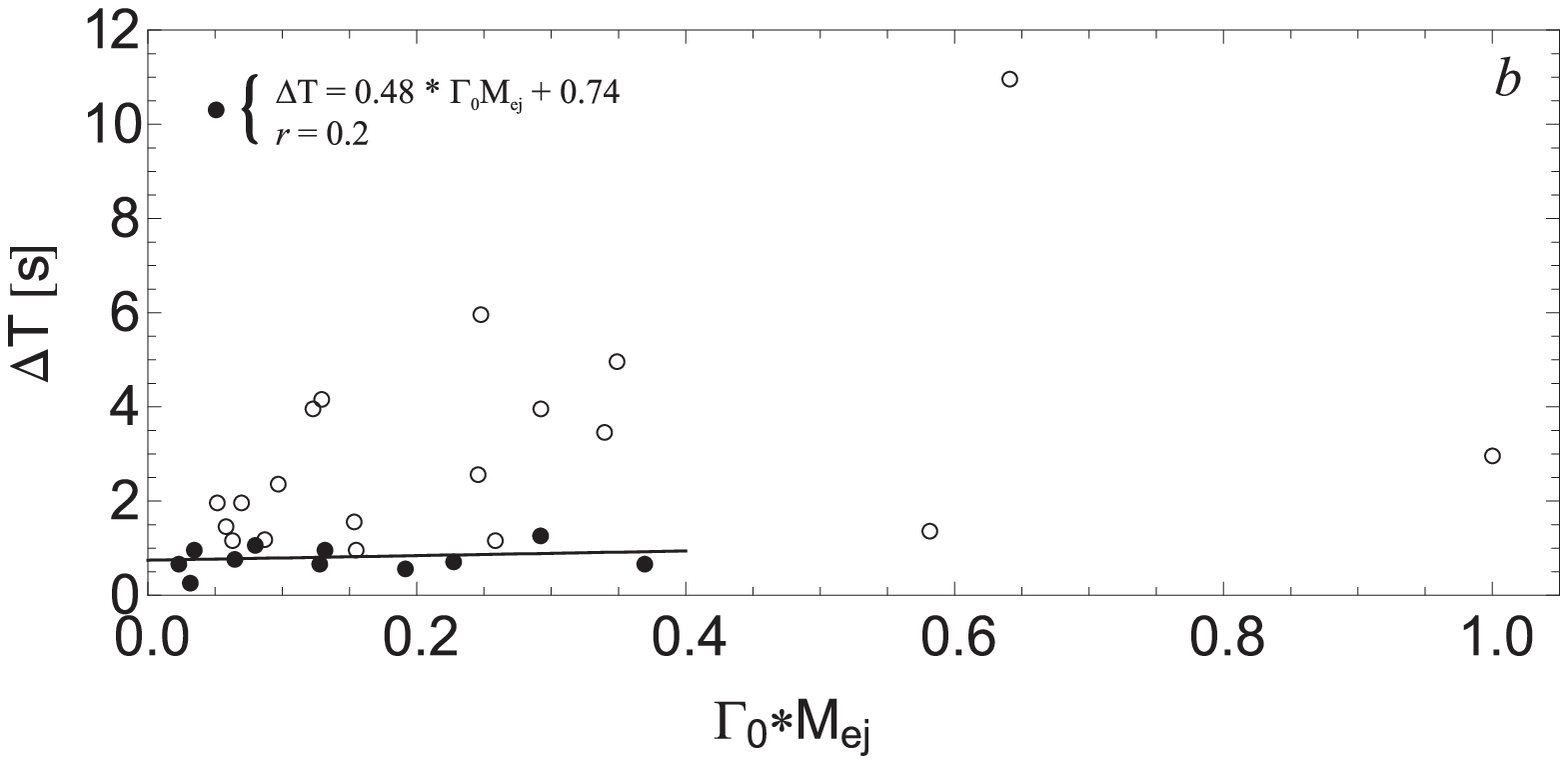}
\includegraphics[width=6cm]{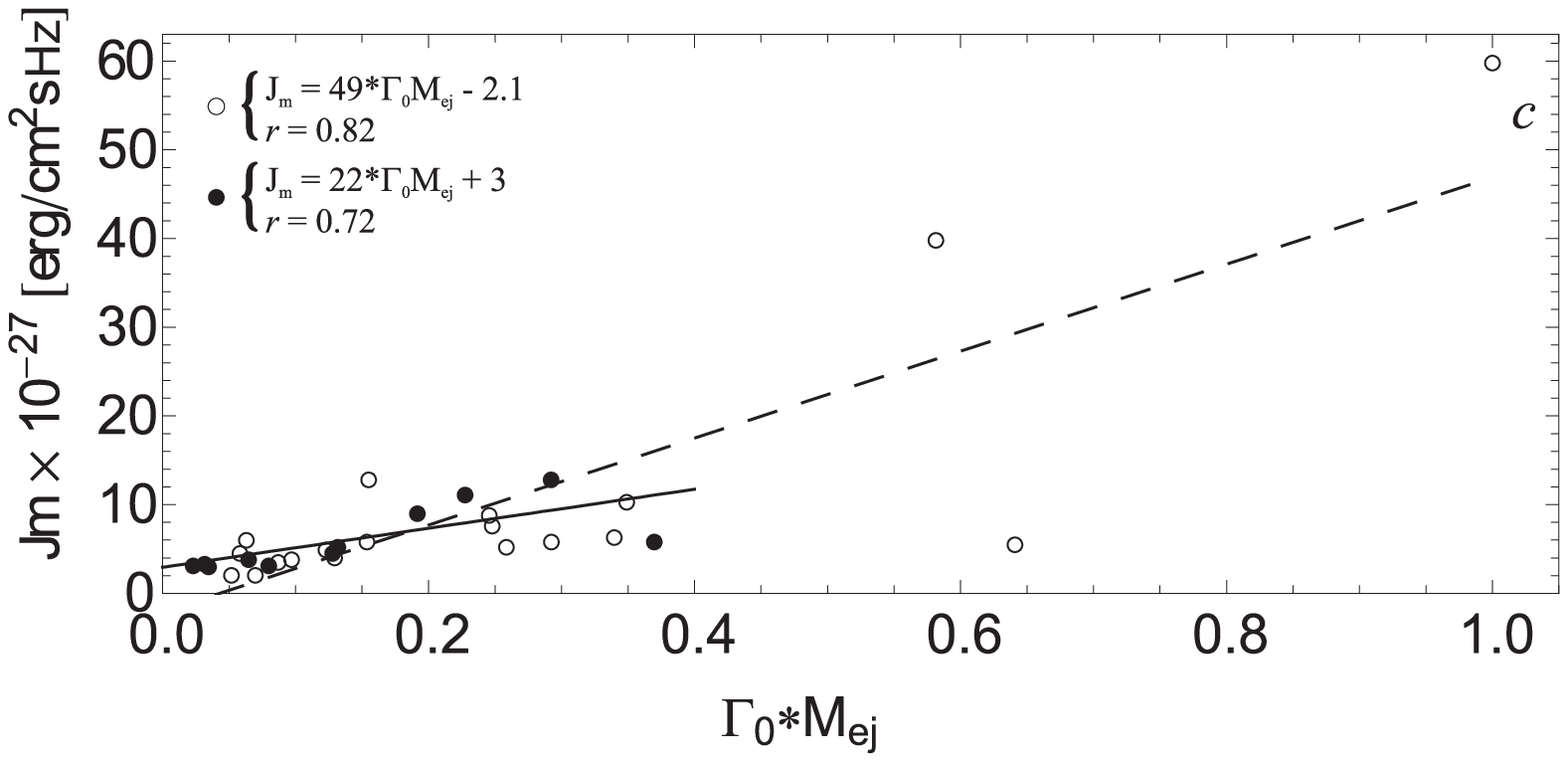}
\includegraphics[width=6cm]{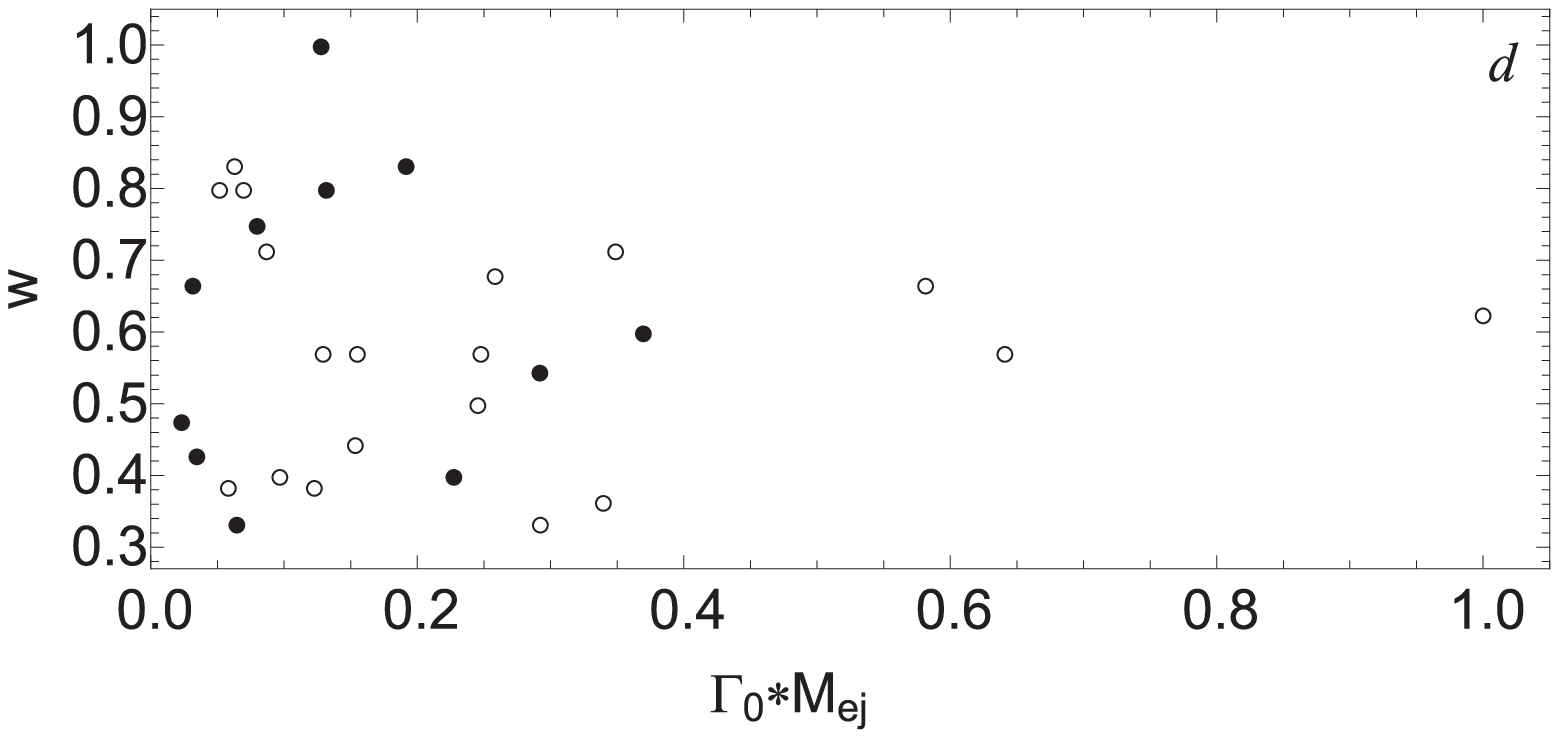}
\caption{Parameter dependance: FWHM, $\Delta t$, $J_m$ and $w$ vs. $\Gamma_0 \cdot M_{\rm ej}$ (scaled to the Max[$\Gamma_0 \cdot M_{\rm ej}$]), panels a - d, respectively. The notation for short and long pulses is same as in Fig. 4 and 5.}
\label{fig06}
\end{figure*}

Additionally, we explore  possible correlations between parameters obtained from fitting the light curves and measured ones (given in \S 3.1). In  Fig. \ref{fig06} we plot measured values against $\Gamma_0 \cdot M_{\rm ej}$, which is proportional to the energy of the incoming shell.

It is interesting that the pulse intensity for small energies ($\Gamma_0 \cdot M_{\rm ej}<$0.2) shows nearly linear trend with the energy (see Fig. \ref{fig06}c), but for  $\Gamma_0 \cdot M_{\rm ej}>0.2$ this trend is not present.

\begin{figure*}
\centering
\includegraphics[width=6cm]{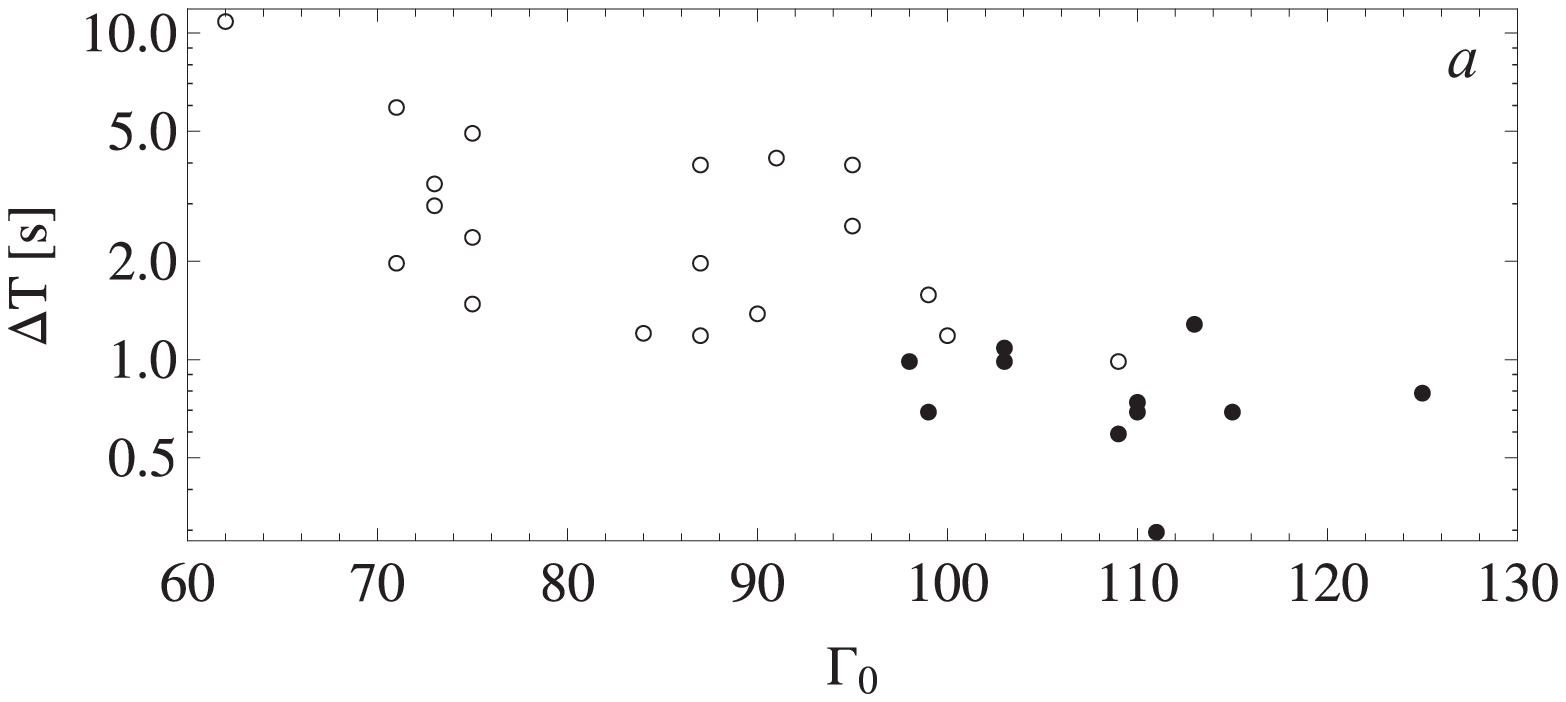}
\includegraphics[width=6cm]{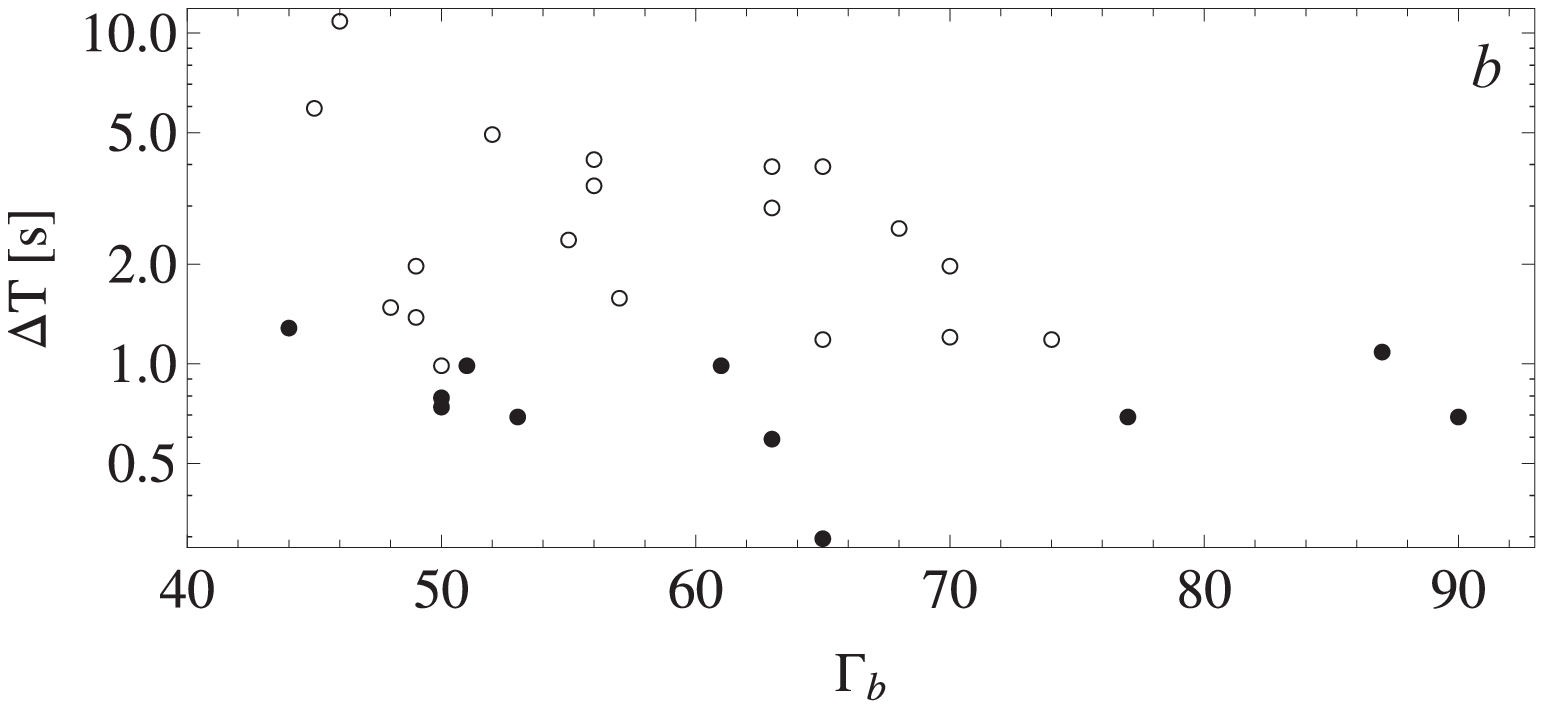}
\includegraphics[width=6cm]{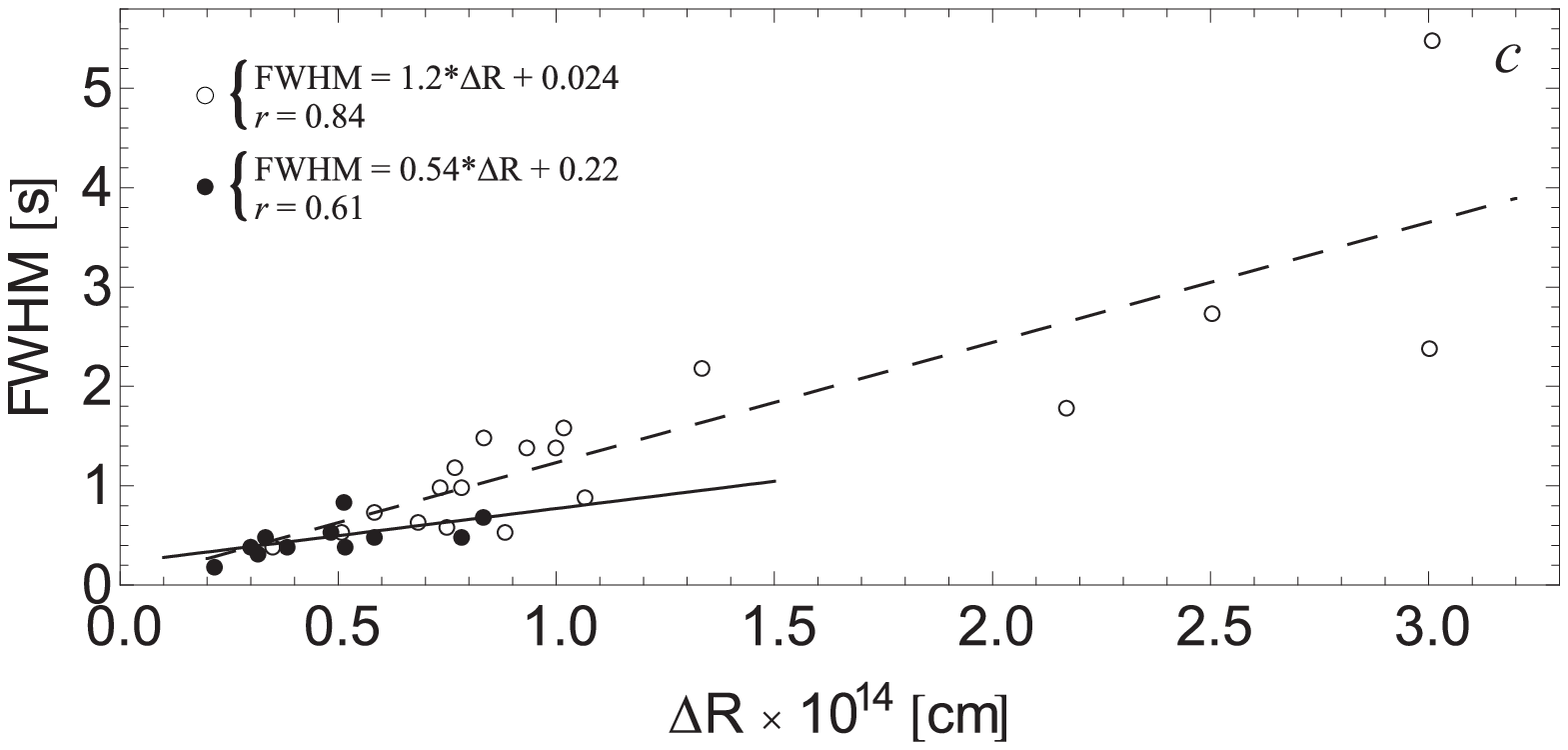}
\caption{Correlation of pulse and shell parameters. Case of $\Delta t$ vs. $\Gamma_0$, $\Gamma_b$ (panels a, b), and FWHM vs. $\Delta R$, (panel c). The notation for short and long pulses is same as in Fig. 4 and 5.}
\label{fig07}
\end{figure*}

In Figs \ref{fig07}a and b we present $\Delta t$ vs. $\Gamma_0$, $\Gamma_b$, respectively and in Fig. \ref{fig07}c FWHM vs $\Delta R$. It is obvious that for a faster barrier and shell the interaction will be shorter. Also, for a broader barrier we obtain long lasting (wider) pulses as it is shown in Fig. \ref{fig07}c. There is a correlation between FWHM and $\Delta R$, with correlation coefficients r = 0.61 for short and r = 0.84 for long lasting GRBs. A linear relationship between FWHM and $\Delta R$ is present as FWHM=a$\cdot \Delta R$ + b (where a = 0.54 and b = 0.22 for short and a = 1.2 and b = 0.024 for long lasting GRBs, see Fig. \ref{fig07}c).

\begin{figure*}
\centering
\includegraphics[width=6cm]{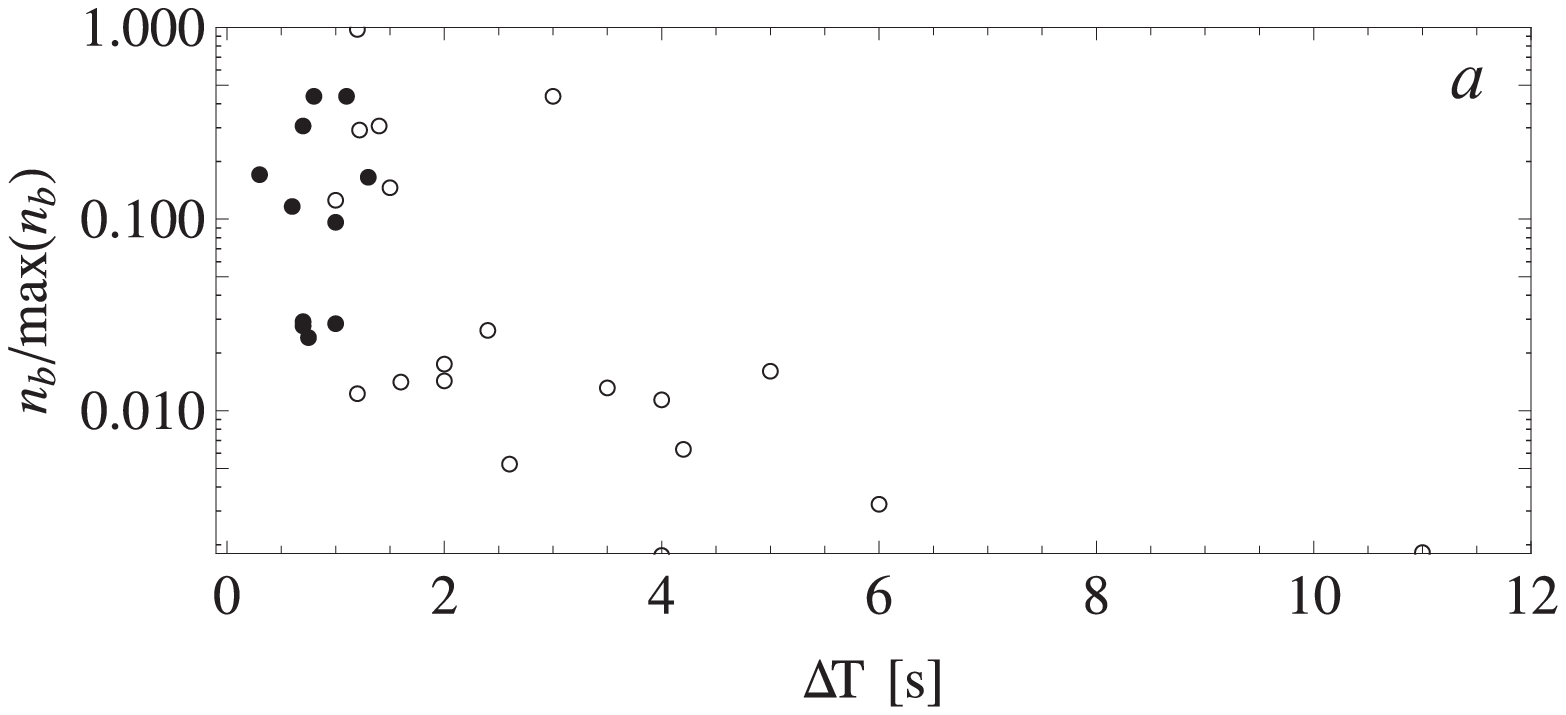}
\includegraphics[width=6cm]{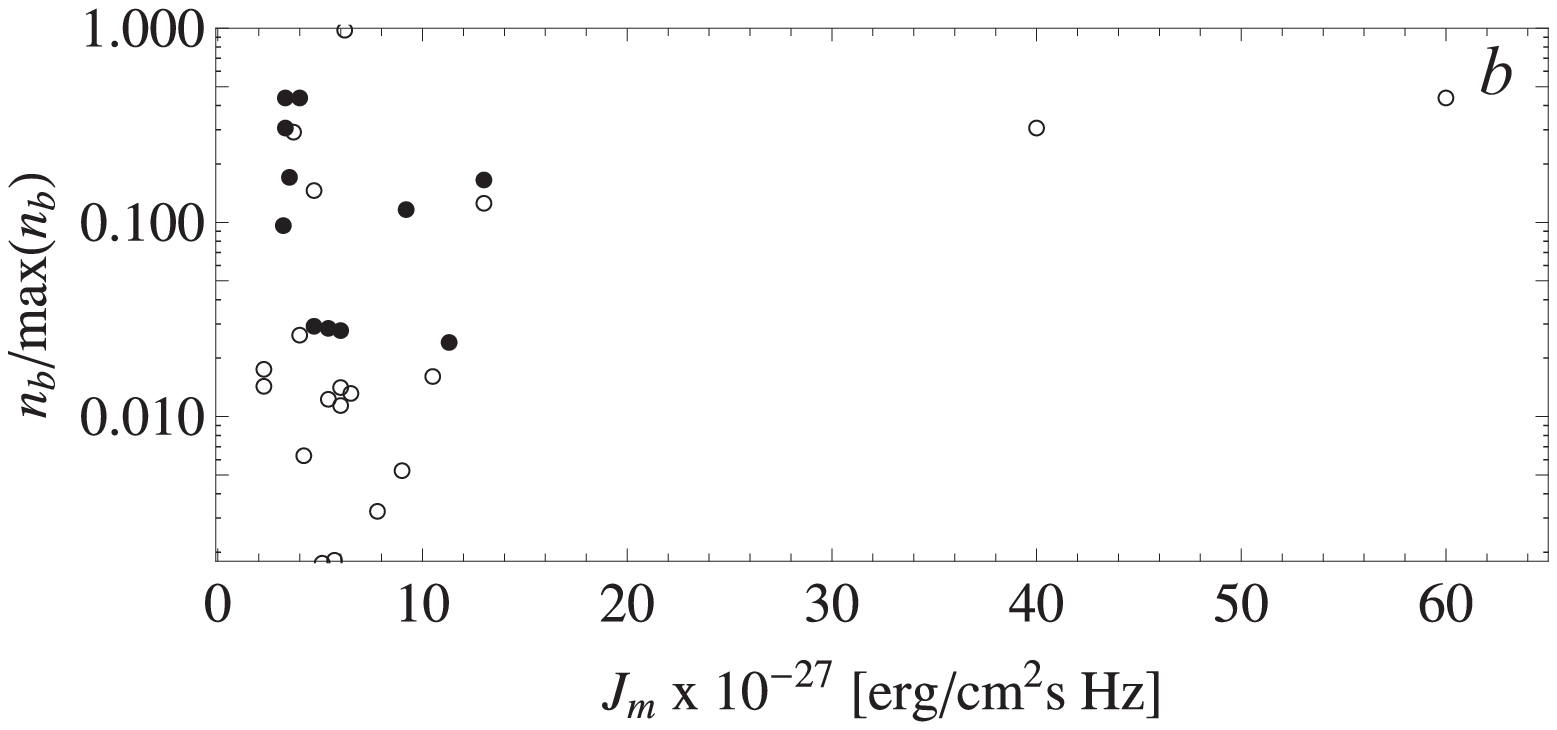}
\caption{Correlation of pulse and shell parameters. Case of $n_b$ on $\Delta T$ (a) and $n_b$ on $J_m$ (b).}
\label{fig08}
\end{figure*}

We found some connections between  $n_b$ vs. $\Delta T$ and $n_b$ vs. $J_m$, as presented in
Fig. \ref{fig08}. This is expected, since for a barrier with higher density there are more intense pulses (Fig. \ref{fig08}a). The smaller values of density produce broader light curve pulses (Fig. \ref{fig08}b). It can be seen from Fig. \ref{fig08}a that short pulses have mainly smaller densities.   This can be easily understood analyzing the physical processes in the moment of collision. When the particle number density is higher more particles are taking part in the interaction, producing a higher gamma-ray flux. But if the $n_b$ decreases the interaction is prolonged, caused by the stretching of the barrier material, so long lasting pulses are produced.

\section{Conclusions}

In this paper we extracted the basic parameters of internal
shock waves during the first phase of a GRB event by fitting 30 observed GRBs (from
BATSE database).  To fit the observed GRB light curves we used the modified internal shock model given in [\refcite{Simic07}], assuming the collision of a fast shell with a slow moving (with respect to the velocity of the shell) barrier. We analyzed the obtained parameters in order to find physical processes behind the GRB origin, and we came to the following conclusions:

(i) Relativistic shell parameters obtained from the fitting of GRB light curves are in a good agreement with expected ones and also with estimations given earlier by other authors mentioned throughout the text.

(ii) The obtained values of internal shell physical parameters for  GRBs with different light curves are in
the short interval, showing that the physical processes behind the GRB creation are
similar, i.e. there should be the ejected mass that collides with surrounding regions - or accumulated slow moving material.

Also, we analyzed possible connections between parameters obtained from the best fitting of GRB light curves with measured ones. From this analysis we can conclude:

(i) There is no correlation between parameters obtained from the best fitting (Figs 4 and 5), only some indication that long GRBs have higher values of Lorentz factor, and we found a slight trend between Lorentz factor of the shell and moving barrier for short pulses.

(ii) There is a correlation between the intensity of pulses and the energy density of the shell only for a low energy pulses ($\Gamma_0\cdot M_{\rm ej}<0.2$, see Fig. 6c).

(iii) The FWHM of GRB light curve pulses is in the correlation with the width of the barrier. Using this we give a
relation between FWHM (that can be measured from observed light curves) and $\Delta R$ that is a parameter of the model (see Fig. 7b).

Finally, we can conclude that the modified internal shock model that assumes also
the barrier density distribution can well describe the first phase of the GRB origin. Moreover, the obtained
parameters for the internal shocks and barriers well fit the physics of the GRBs.

\section*{Acknowledgments}

The work was supported by the Ministry
of Education and Science  of R. Serbia through the
project "Astrophysical Spectroscopy of Extragalactic Objects".
We would like to thank to the anonymous referee for very useful comments and
suggestions and D. Ili\'c for careful reading of this manuscript and comments.

\end{document}